%% file: mainSRDD.tex
\documentclass[11pt]{article}

\usepackage{ML,TZ}
\usepackage{appendix}
\usepackage{float}
\usepackage{dcolumn}
\usepackage{adjustbox, caption, subcaption}
\usepackage{parskip}
\def\fullcite#1{\cite{#1}}

\usepackage{setspace}
\usepackage{caption}

\def\commentnj#1{\color{black}#1 \color{black} }
\begin{document}

\title{Longitudinal Position and Cancer Risk in the United States Revisited}

\author[1]{Jin Niu}

\author[2]{Charlotte Brown}

\author[3]{Michael Law} 
\author[4]{Justin Colacino}
\author[5]{Ya'acov Ritov }

\affil[1]{Department of Economics, Brown University.}
\affil[2]{Department of Statistics, University of Michigan, Ann Arbor}
\affil[3]{Seminar für Statistik, ETH Zürich}
\affil[4]{School of Public Health, University of Michigan, Ann Arbor}
\affil[5]{Department of Statistics, University of Michigan, Ann Arbor}

\maketitle


\vspace{0.5cm}
\textit{Running title}: Longitudinal Position and Cancer Risk in the U.S. Revisited


\vspace{0.5cm}

\vspace{0.5cm}
\textbf{Corresponding Author:}

Jin Niu

Brown University

69 Brown St. Box B, 

Providence, RI, 02912-9091, United States

Phone: 925-818-6953

Email: jin\_niu@brown.edu

\vspace{0.5cm}
\textbf{Other Contacts:}

Charlotte Brown (cabrownn@umich.edu)

Michael Law (michael.law@stat.math.ethz.ch)

Justin Colacino (colacino@umich.edu)

Ya'acov Ritov (yritov@umich.edu)

\vspace{0.5cm}
\textbf{Conflict of interest statement:} The authors declare no potential conflicts of interest.

\newpage
\begin{abstract}
    \input{Sections/0_Abstract}
\end{abstract}

\newpage
\section{Introduction}
\input{Sections/1_Introduction}

\section{Materials and Methods}
\input{Sections/2_Data}

\input{Sections/3_Methods}

\section{Results}
\input{Sections/4_Results}

\section{Discussion}
\input{Sections/5_Discussion_Conclusion}

\section{Acknowledgements}
\input{Sections/Acknowledgements}

\bibliographystyle{ieeetr}
\bibliography{ref}

\clearpage
\section{Tables}
\input{Sections/Tables}

\clearpage
\section{Figure Legends}
\input{Sections/Figure_Legend}

\clearpage
\section{Figures}
\input{Sections/Figures}

\clearpage
\appendix 
\section{Supplementary Information}
\setcounter{figure}{0}
\setcounter{table}{0}
\input{Sections/supplementary}

\end{document}

%% file: Sections/0_Abstract.tex

\textbf{Background}: The debate over daylight saving time has surged, with interests in the effects of sunlight exposure on health. \commentnj{Prior studies simulated daylight saving time and standard time conditions by analyzing different locations within time zones and neighboring areas across time zone borders.} 

\textbf{Methods}: \commentnj{We analyzed cancer incidence rates from various longitudinal positions within time zones and at time zone borders in the contiguous United States. Using data from State Cancer Profiles (2016-2020), we analyzed total cancer of 19 types and specific rates for eight cancers, adjusted for age and includes all demographics. Log-linear regression is used to replicate a previous study, and spatial regression models are employed to explore discontinuities at borders.} 

\textbf{Results}: \commentnj{Cancer rate differences lack statistical significance within time zones and near borders for total cancer and most individual cancers. Exceptions included breast, prostate, and liver \& bile duct cancers, which exhibited significant relationships with relative position at the 95\% significance level. Breast and liver \& bile duct cancers saw decreases, while prostate cancer incidence increased from west to east within time zones.} 

\textbf{Conclusions}: \commentnj{Relative position does not have a significant impact on cancer incidence, hence cancer development in general. Isolated exceptions may warrant further investigation as more data becomes available.}

\textbf{Impact}: \commentnj{Our findings challenge prior research, revealing numerous inconsistencies. These disparities urge a reconsideration of the potential disparities in human health associated with daylight saving time and standard time. They offer insights contribute to the ongoing discussion surrounding the retention or abandonment of DST.}


%% file: Sections/1_Introduction.tex
Gu et al. \cite{gu2017longitude} investigated the relationship between the total cancer rate of 23 types, and subgroups of breast cancer, with the relative position in a time zone (the difference between a county's longitude and the standard meridian of the time zone it is in) in the continental United States. They found that risk for developing cancer is increasing from the east to the west within a time zone, which may be attributed to a greater lag between social time and solar time experienced by western time zone residents. However, their study was limited to 607 counties in 11 states, representing only about 20\% of US counties, and only included data for the white race. This paper aims to expand on their analysis using total cancer and eight specific types of cancer incidence data, with the largest observation including 2853 counties from 48 states. We explore the relationship of cancer rate with respect to longitudinal position in a time zone using a linear model and a spatial discontinuity framework with natural splines. \commentnj{The statistical method we chose to employ enhances accuracy and robustness of our analysis, and enables us to better capture the relationship within time zones as well as the potential discontinuity of cancer rate around borders.}

\commentnj{Recently, there has been increased attention on the potential for circadian rhythm disruption to impact the development of multiple cancer types. For example, the International Agency for Research on Cancer (IARC) has identified that shift-work involving circadian disruption is “potentially carcinogenic to humans (Group 2A)” \cite{international2010painting}, with potential links to breast, colorectal, lung, prostate, and skin cancers. However, whether the effects of more subtle alterations in circadian patterns at the population level is a cancer risk factor is still unsettled. Recent work has examined the discrepancy between an individual’s biological time and their “social time” based on their longitudinal location in a time zone, showing an increase in circadian misalignment as one is located further west in a time zone. Moreover, there has been a debate about whether the nation should keep the practice of switching between standard time and daylight saving time, and if not, which is preferred to keep. The main argument in favor of standard time is based on the long-term health effects of starting the day an hour later (as an hour of sunlight gets allocated to the evening from the morning when switching to standard time). This is based on studies that compared the effect of longitudinal position on cancer epidemiology. For example, a county just west of a time zone boundary is in standard time as compared to the county just east of the boundary, so examining incidence rates of counties from both sides mimic examining them under standard time and daylight saving time. Thus, there could be broad public health implications relative to time zone position and disease outcomes, including cancer risk, which could be directly addressable by policy changes.}

Several studies have generalized the differences of standard time and daylight saving time to different longitudinal positions within a time zone. When investigating the impact of latitude and position in time zone (longitude) on cancer incidence, cancer mortality, and expected lifespan, Borisenkov \cite{borisenkov2011latitude} used data from the European part of Russia and 31 regions of China. Their findings revealed that as latitude increases and when moving from the eastern to the western border of the time zone, both cancer incidence and cancer mortality exhibit an upward trend, which are in agreement with Gu et al.'s findings \cite{gu2017longitude}. Giuntella and Mazzonna \cite{giuntella2019sunset} focused on time zone boundaries where the sharp discontinuity in the timing of sunlight with respect to social time occur. They found respondents residing on the late sunset side of the border (western side of a time zone, eastern side of the border) exhibited a pattern of shorter sleep duration and poorer sleep quality. Moreover, these respondents are also more likely to be overweight and have less desirable health outcomes. These health outcomes are consolidated into a composite index encompassing obesity, diabetes, cardiovascular diseases, and specific types of cancer including breast, colorectal, and prostate cancers. 

In this paper, we revisit the relationship of cancer rate and relative position within the time zone and around time zone borders. We employ a linear model to regress cancer incidence rate on relative position, and a spatial discontinuity framework with natural splines to examine the potential discontinuity in cancer rate around the time zone borders. Our study focuses on the four time zones of the contiguous United States: Eastern, Central, Mountain, and Pacific, as shown in Figure \ref{fig:timezone_map} in the supplementary section.
We acknowledge that the time zones within the United States are not created uniformly. The timezone borders are intended to fall every 15 degrees longitude starting with the Eastern time zone (UTC -5) at meridian $75^\circ$ west, but the division of counties to time zones follow the county borders more than the rigid 15-degree cutoff, with a number of states and a few counties observing different time zones. 




%% file: Sections/2_Data.tex



\subsection{Data}

Cancer incidence rate by county is calculated as the number of cases per 100,000 population averaged over the period from 2016 to 2020. It covers 19 types of cancer of all stages (the complete list of cancers is listed in Table \ref{cancertype} in the supplementary section), for all races, both sexes, and adjusted for age. \commentnj{In addition, according to the International Agency for Research on Cancer (IARC) \cite{international2010painting}, we include four types of hormonally associated cancers that are potentially more closely related to circadian disruption caused by timezone change. They are breast, ovary, prostate, and thyroid cancer. Besides, we also include four most prevalent cancers in the U.S., ranked by number of cases and deaths, for comparison. They are liver \& bile duct, lung \& bronchus, colon \& rectum, and pancreas cancer.}

Data for total cancer incidence rate include 2853 available observations (Figure \ref{fig:timezone_map}, no data in dark areas). 261 counties are missing or excluded for the following reasons: 

\begin{itemize}

    \item Alaska and Hawaii (33 counties) for they comprise distinct time zones from the 48 contiguous states;
    
    \item Arizona (15 counties) for most of Arizona does not observe daylight saving time, and is effectively in Mountain Standard Time and Pacific Standard Time, each for half of the year;
    
    \item Kansas, Minnesota, and others (203 counties) with missing cancer incidence rates;
    
    \item Counties split between two time zones (6 counties: ND: Dunn County, McKenzie County, Sioux County. SD: Stanley County. NE: Cherry County. FL: Gulf County);
    
    \item Counties that are outliers (3 counties: Ziebach County, South Dakota; Union County, Florida; Richmond County, Virginia) that either has unusually high or unusually low rates;
    
    \item \commentnj{Washington D.C. (1 county/city) is excluded as it does not belong to any state and doesn't have the state level variable to control for.}
    
\end{itemize}

\commentnj{For the specific types of cancers, we have 2615 observations for breast cancer, 890 for ovary cancer, 2623 for prostate cancer, 1191 for thyroid cancer, 1117 for liver \& bile duct cancer, 2441 for lung \& bronchus cancer, 1955 for colon \& rectum cancer, and 1090 for pancreas cancers. Besides missing data, we also excluded the aforementioned counties from datasets to maintain consistency.
}

Following Gu et al. \cite{gu2017longitude}, we control for latitude, median income per capita, and smoking rate (percent of individuals that are smokers) at the county level. In addition, we adjust for race (classified into six main categories: white, black, native, Asian, Hispanic, and others), educational attainment (categorized as below high school, high school, some college, and college and above), access to healthcare (measured as per capita number of medical doctors), obesity rate (percentage of adults with obesity), air pollution (average daily PM2.5 concentration), and water violation (indicator of presence of health-related drinking water violations, gathered from Safe Drinking Water Information System). 

We incorporate elevation as an additional control variable. According to Giuntella and Mazzonna \cite{giuntella2019sunset}, higher altitudes are associated with increased exposure to sunlight, which is potentially related to cancer development. Specifically, the sun rises earlier and sets later in higher altitudes. Since high elevation is more common in the western part of the United States (in the Mountain Time and Pacific Time zones), including elevation accounts for potentially unobserved variation in cancer rate in these regions.

To account for socioeconomic differences, we retrieved median income per capita data from the 2020 US census. It is worth noting that Virginia, different from other states, has 38 independent cities aside from 95 counties, each having its own Federal Information Processing Standards (FIPS) code. The US census combines independent cities with their surrounding counties when reporting median income. To ensure consistency in our analysis, we calculated a population-weighted average of all other variables for these combined cities and counties to match the level of observation in the median income data set. The complete list of counties and cities that were combined in Virginia is listed in Table \ref{virginia_append} in the supplementary section.

The main explanatory variable in our analysis is longitude, given as the population center of each county provided by the US Census Bureau. In addition, we include the relative longitudinal position (PTZ) following Gu et al. \cite{gu2017longitude}. It is calculated as the difference between county's longitudinal position and the standard meridian of their respective time zones, which are -75, -90, -105, and -120 for Eastern, Central, Mountain, and Pacific, respectively. PTZ as a continuous variable is presented in Figure \ref{fig:rel_position}.

We present summary statistics of total cancer rate, relative position, and various control variables in Table \ref{sumstats} in the supplementary section. The average PTZ is around -4, indicating the majority of counties are located towards the western side of the time zone. The mean elevation is 393 meters, with a maximum elevation of 3163 meters observed in the Mountain/Pacific time zone. As for access to health care, the average number of medical doctor per capita is 0.001, with the maximum being 0.038 and the minimum being less than 0.0005. The table also shows statistics for other control variables, including educational attainment, median income, obesity rate, smoking rate, air pollution, water violations, and race composition. 

Figure \ref{fig:cancerrate_map} displays counties with their respective total cancer rates. The analysis on specific types of cancers have less observations compared to the total cancer dataset. Figure \ref{fig:hormon_cancer_ci_map} and Figure \ref{fig:oth_cancer_ci_map} in the supplementary section display similar maps of cancer rates by county.



\subsection{Data Availability Statement}

The cancer incidence data analyzed in this study were obtained from the National Cancer Institute and Centers for Disease Control and Prevention at State Cancer Profiles.

The data that control for geographic, socioeconomic, and health-related variations are collected from the Bureau of Economic Analysis, the US Department of Agriculture, Health Resources \& Services Administration, and the US Census Bureau. The complete correspondence for data and sources is given in Table \ref{datasources} in the supplementary section.


%% file: Sections/3_Methods.tex

\subsection{Methods}
To investigate the relationship between cancer rates and relative position (PTZ), we begin by replicating Gu et al.'s method \cite{gu2017longitude}, and regress the natural logarithm of cancer incidence rate, which they refer to as covariate-adjusted rate ratios (RR), on PTZ using a weighted least squares linear regression. We use a least squares linear regression model and control for variables listed in Equation \ref{semiparametric}. We first apply this model using the 607 counties included in Gu et al.'s analysis \cite{gu2017longitude} with total cancer rate data. Then, we extend the model to include all 2853 counties available to us. \commentnj{For a significant relationship to be detected that aligns with Gu et al.'s findings, the resulting coefficient estimate must be negative and significant at the 0.01 level (having a p-value less than 1\%). It is worth noting that using PTZ assumes the uniform distribution of cancer rates across all four time zones.}

Then, to avoid making such strong assumption or enforce periodicity across time zones, we employ a semiparametric model for cancer incidence rate that is given by 
\begin{equation}
\label{semiparametric}
\begin{split}
    \text{Cancer Incidence (RR)} = & f(\text{Longitude} \times \text{Time Zone}) + \beta_{1} \text{State} + \beta_2 \text{Latitude} \\
    & + \beta_3 \text{Race} + \beta_4 \text{Education} + \beta_5 \text{Medical Doctors} \\
    & + \beta_6 \text{Income} + \beta_7 \text{Obesity} + \beta_8 \text{Smoking} + \beta_9 \text{Air Pollution} \\
    & + \beta_{10} \text{Water Violation} + \beta_{11} \text{Elevation} + \epsilon,
\end{split}
\end{equation}
where $f(\cdot)$ is assumed to be a smooth nonparametric function on longitudes instead of PTZ. We are interested in the effect of $f(\cdot)$ near the boundary of two time zones (Eastern and Central, Central and Mountain, Mountain and Pacific). \commentnj{Our primary focus centers on assessing whether the estimated cancer incidence rate (represented by the solid lines) within one time zone significantly differs from that within its neighboring time zone. To determine this, we examine whether the resulting incidence estimate falls outside the 95\% confidence band of the neighboring incidence estimate.  If such scenario occurs, we infer the presence of a discontinuity in cancer incidence at the border of these two time zones. If this discontinuity is consistently observed across multiple border regions, it becomes reasonable to argue that the disparity in cancer incidence is associated with disruptions in the timing of natural light, which in turn can serve as a stepping stone for exploring the long-term health implications of daylight saving time (DST) and standard time.}

We use natural cubic splines with 6 degrees of freedom (chosen by cross-validation) on longitudes to model cancer rate. The \texttt{ns()} function in \texttt{R} is used to generate basis matrix for piece-wise cubic splines with specified interior knots and natural boundary conditions \cite{r_ns}. We specify degrees of freedom instead of a sequence of interior knots when using splines. Our model considers a nonparametric functional form of longitude on both sides of the boundary and within the entire time zone. It is a generalization of the regression discontinuity framework used by Giuntella and Mazzonna \cite{giuntella2019sunset} who only focused around the border. We construct the 95\% confidence interval by performing 1000 bootstrap replications for each model, normalized by dividing by the square of standard error of cancer incidence rate.

We subsequently apply the same procedure, fitting a linear model and regression discontinuity framework with natural splines, is applied to the specific types of cancer: breast, ovary, prostate, thyroid,  liver \& bile duct, lung \& bronchus, colon \& rectum, and pancreas, to examine the specific relationships between the cancer incidence and longitudes as well as cancer incidence and PTZ.

%% file: Sections/4_Results.tex
\subsection{Replication of Relative Position (PTZ)}\label{repli_result}

The resulting linear estimates, accompanied by their corresponding 95\% confidence intervals, on relative position (denoted as PTZ) are shown in Figures \ref{fig:linear_approx}, including a total of 607 and 2853 counties respectively. Figure \ref{fig:607linear} presents the output derived from using the same methodology and counties as used by Gu et al., while Figure \ref{fig:alllinear} presents the output resulting from employing the same method with a broader set of observations. The x-axis is PTZ, the relative longitude in a time zone, and the y-axis is total cancer rate (standardized by the median PTZ data point). Both regression lines demonstrate negative relationships between cancer incidence and PTZ, suggesting that the regions in the western end of time zones tend to exhibit higher cancer rates in comparison to their eastern counterparts. However, this correlation is only statistically significant (at the 95\% level) with the data set containing only 607 counties; the relationship becomes no longer significant as the data set includes more observations. 

While the finding given in Figure \ref{fig:607linear} is consistent with the conclusion made by Gu et al. \cite{gu2017longitude}, the inconsistency of results from differing data set sizes suggests that the smaller data set, selected by Gu et al., is likely biased and not representative of the U.S. counties in general. To elucidate this, we generated a relative weight plot of the 607 counties according to their PTZ and color-coded by their respective time zones. As presented in Figure \ref{fig:rel_weights} in the supplement section, this visualization reveals a non-uniform distribution of counties from the four time zones along the PTZ spectrum. Specifically, counties from the Eastern time zone mainly populate the lower end of PTZ values (westward in the time zones), while counties from the Pacific time zone dominate the higher end of PTZ values (eastward in the time zones). This signifies that higher cancer rates observed in the Eastern time zone likely contribute to the higher estimated cancer rates at the western extremities of time zones. Conversely, the lower cancer rates found in the Pacific time zone potentially led to the lower estimated cancer rates at the eastern extremities of time zones. Consequently, the apparent significant negative relationship between cancer rates and PTZ, as identified by Gu et al. in Figure \ref{fig:607linear}, is likely influenced by the uneven distribution of counties from different time zones in their data set.

Figure \ref{fig:linear_boot} in the supplement section illustrates the linear approximation of cancer rate by longitude in each time zone with the 95\% bootstrap confidence band. The plot highlights the irregularity of the total cancer rate trend with respect to PTZ within different time zones. Specifically, only the Eastern time zone exhibits a statistically upward trend of cancer rate with longitude, setting it apart from the other time zones. At the same time, Mountain and Pacific time zones present wide confidence bands, which is likely attributed to the fewer units (counties) included in these time zones.

Table \ref{linearmod_relpos} displays the estimated coefficients and their corresponding 95\% confidence intervals using the data set of all 2853 counties. The coefficient for relative position is estimated to be around -0.4, but is not statistically significant with a confidence interval including 0 ($[-0.898, 0.098]$). This result is in agreement with Figure \ref{fig:alllinear}, indicating no significant relationship between cancer incidence rate and PTZ is present. \commentnj{A complete comparison between our findings and Gu et al.'s findings is presented in Supplementary Table \ref{sum_gu_results}.}

\subsection{Our Results}

We note that the geography of the United States' east coast is directed northeast-southwest: as we move eastward, we also move northward. As a result, New England represents the entire longitudinal region it is in by being the only observations, which means that high cancer incidence in New England result in high estimate of cancer incidence of the east coast (the eastern end of the Eastern time zone). This explains the result we obtained later for total cancer incidence, presented in Section \ref{sec:total}.

\subsubsection{\commentnj{Hormonally Associated Cancers}}

\commentnj{Among cancers that are linked to hormones---namely breast, endometrial, ovary, prostate, testis, and thyroid cancer---we were only able to collect data for breast, ovary, prostate, and thyroid cancer. Hence, our analysis focused on these four cancer types to stand as representative examples of hormonally associated cancers.}

\commentnj{The linear approximation results are presented in Table \ref{linmod_relpos_hormon_cancer} within the supplementary section. The visual presentation is given in Figure \ref{fig:hormon_lin} in the supplementary section. As both the Table and Figure show, breast cancer rate demonstrate significant decrease, whereas prostate and thyroid cancer rates exhibit significant increases when moving from west to east within a time zone. The relationships are statistically significant at the 0.01 significance level. More specifically, for each longitudinal degree eastward, the incidence for breast cancer decreases by 0.285 on average. Similarly, for each longitudinal degree eastward, prostate and thyroid cancer incidences increase by an average of 0.387 and 0.123, respectively. This corresponds to RR per five degrees of longitude toward the east (equivalent to 20 mins advance of sunrise, as noted by Gu et al.) being -1.425 for breast cancer, 1.935 for prostate cancer, and 0.615 for thyroid cancer. Our result for breast cancer is consistent with that of Gu et al.'s, as both are significant and demonstrate a decreasing trend in incidence rate when moving from west to east. However, results for prostate and thyroid cancers contradict with what Gu et al. had obtained by presenting opposite signs for the relationship.}

\commentnj{The inconsistent signs of relationships of the four hormonally associated cancer incidences with relative position do not support that position in a time zone influence development of hormonally associated cancers.}

\commentnj{Moreover, it is worth noting that graphs in supplementary Figure \ref{fig:hormon_lin} use bootstrap to generate the confidence interval, which bears some level of randomness. Meanwhile, the intervals in Table \ref{linmod_relpos_hormon_cancer} (supplement) are derived from $t$-tests performed in linear regression. This difference in methodology resulted in 95\% confidence bands in Figure \ref{ps_lin} and Figure \ref{thy_lin} seemingly covering the horizontal axis when the linear regression output suggests that the 95\% confidence intervals do not cover 0. It's vital to emphasize that this does not represent a contradiction, but rather arises from the application of distinct methods.}

\commentnj{Further, we investigate cancer incidence discontinuity around time zone borders using natural splines. Estimated cancer incidence by longitude, accompanied by 95\% confidence bands, is presented in Figure \ref{fig:hormon_boot}. Among them, only prostate cancer exhibit discontinuity at the border between the Eastern and Central time zones (Figure \ref{ps_spl}). However, this discontinuity is not consistently present for prostate cancer at other time zone borders, neither is it present for other hormonally associated cancers. \commentnj{As a result, we do not conclude that discontinuity in time zone is related to cancer incidence.}}

\commentnj{Lastly, thyroid cancer exhibits high incidence in the eastern region of the Eastern time zone, which is the northeast region of the U.S. Thyroid cancer therefore could be responsible for the high estimated total cancer incidence observed in the east coast that is shown in Figure \ref{fig:comp_spl}.}

\subsubsection{\commentnj{Most Prevalent Cancers}}

\commentnj{Performing linear approximation on the four most prevalent types of cancer, namely liver \& bile duct, lung \& bronchus, colon \& rectum, and pancreas cancer, on relative position has resulted in output displayed in Table \ref{linmod_relpos_subcancer} within the supplementary section. Their respective linear estimates with corresponding 95\% confidence intervals are presented in Figure \ref{fig:subcancer_lin} in the supplementary section. Because the data set for liver \& bile duct cancer has much fewer observations (1117), of which only 54 counties belong to the Mountain time zone, we drop the observations from the Mountain time zone to preclude over-representation they might exert on the result.
}
\commentnj{At the 0.01 significance level, the coefficient estimate is significant for liver \& bile duct cancer after excluding the Mountain time zone observations (column 3 of Table \ref{linmod_relpos_subcancer}), and exhibits a negative relationship. Specifically, for each longitudinal degree eastward, the incidence for liver \& bile duct cancer decreases by
an average of 0.102, which corresponds to RR per five degrees of longitude eastward being -0.510. This is consistent with Gu et al.'s finding on liver \& bile duct cancer in sign and significance, despite our estimated correlation (coefficient) is smaller in magnitude. The estimate for lung \& bronchus cancer rate has a weaker significance at the 0.1 level and exhibits a positive relationship. Neither of the coefficient estimates of colon \& rectum and pancreas cancer reaches statistical significance. }

\commentnj{Figure \ref{fig:subcancer_boot} presents the estimated cancer incidence rates by longitude and time zone using natural splines. Focusing on time zone borders, only colon \& rectum cancer exhibit discontinuity in incidence rate around the Eastern and Central time zone border (Figure \ref{col_spl}). The estimated incidence rate at the western end of Eastern time zone deviates from its neighboring counterpart in the Central time zone, exhibiting lower values for colon \& rectum cancer incidence. However, this region is also accompanied by wider confidence band. This explains that the estimates are greatly influenced by the few observations within that specific longitudinal region (the upper peninsula of Michigan). No significant discontinuity around the border is present for other cancers.
}

\subsubsection{Total Cancer}\label{sec:total}

Lastly, natural splines is performed on total cancer rate. Figure \ref{fig:comp_spl} presents the estimation. Corresponding to Giuntella and Mazzonna's findings \cite{giuntella2019sunset}, we focus on the overlapping region of time zones. Among the three time zone borders, no significant deviation in cancer incidence is observed as incidence estimates (solid lines) do not lie outside of the confidence band of the neighboring line. Since time zone boundaries are drawn irregularly, there are fewer observations around the borders than in the middle of the time zones, which could result in wider confidence bands. 

We created uniform 95\% confidence band for each time zone overlapping region to investigate the relationship around borders in detail. First, we define the overlapping regions---those formed by Eastern and Central, Central and Mountain, and Mountain and Pacific. For the region between Eastern time zone and Central time zone, we take longitudinal region bounded by the east most longitude in Central time zone and the west most longitude in Eastern time zone. The rest regions are determined in a similar fashion. Then, we take the widest 95\% confidence interval within each region, and make it uniform across that region. As shown in Figure \ref{fig:comp_spl_unif}, the lines above and below the estimates are the uniform confidence bands. Again, no solid line lies outside of the confidence band of the adjacent time zone, which is in agreement with the observation that there is no significant difference in cancer incidence around borders of time zones, under 95\% confidence level.

%% file: Sections/5_Discussion_Conclusion.tex
This paper aims to investigate the relationship between cancer incidence rate and longitudinal position, denoted by relative position in a time zone (PTZ) and longitudes. We use PTZ to explore the trend of cancer rates within time zones, and use longitudes to study both the trend and the potential discontinuity at time zone boundaries. We employed both linear regression and a spatial discontinuity framework with natural cubic splines. This analysis is carried out on total cancer rate of 19 types and eight individual cancer rates. 

Upon controlling for latitude, elevation, education, healthcare, income, obesity rate, smoking rate, pollution, and race, the linear model did not reveal a significant relationship between relative position and the total cancer incidence rate. This finding stands in contrast to an earlier research with a smaller data set. Our work suggests that longitudinal position does not exhibit a general correlation with cancer incidence, and the variations in cancer rates for specific types are inconsistent. Moreover, our semi-parametric model using natural splines indicated no significant discontinuities in incidence rates around time zone borders for total cancer and most individual cancers. Exceptions were noted at the Eastern to Central time zone border, where discontinuities were observed for prostate and colon \& rectum cancers. 

Examining neighboring counties situated around time zone borders allows us to simulate conditions similar to a single observation under standard time (on the western side of the border) and daylight saving time (on the eastern side of the border). This distinction translates into an additional hour of morning sunlight for those on the western side and an extra hour of evening sunlight for those on the eastern side. However, our investigation reveals no statistically significant differences in cancer rates either within a time zone (between its two ends) or across the two sides of a border. Consequently, our findings indicate that there is no discernible cancer-related reason to favor one time system over the other.

\commentnj{Studies have also examined the impacts of daylight saving time on mental and physical health. The American Academy of Sleep Medicine has advocated for year-round standard time due to various health concerns \fullcite{rishi2020daylight}. Their statement highlights that DST is less compatible with human circadian rhythm due to late evening exposure to light. Moreover, the one-hour shift of sunlight from morning to evening when transitioning into DST can induce a phase delay, leading to chronic sleep loss, especially for individuals with early morning social obligations such as students. This phenomenon, known as ``social jet lag," refers to the misalignment between social obligations and the innate circadian rhythm, and has been linked to an increased risk of cardiovascular disease \fullcite{MeiraeCruz2019, young2017circadian}. In the western part of a time zone where sunset occurs at a later clock time, social jet lag may be more pronounced compared to its eastern counterpart. However, Heboyan, Stevens, and McCall's \cite{heboyan2019effects} analysis of data from Augusta University Medical Center over a period of approximately 2 years did not find significant association between DST and emergency department visits related to mental and behavioral health conditions. }

Our study is subject to several limitations. Firstly, there are constraints on the generalization of our findings along the coastlines due to the lack of observations at lower latitudes. As explained in the previous section, the United States' east coast follows a northeast-southwest direction, and the west coast is oriented northwest-southeast. This means that moving eastward along the east coast is also a northward shift, and moving westward along the west coast is also a northward progression. This geographic alignment results in a dominance of counties from the northeastern part of the U.S., namely New England, and a similar pattern holds for the western coast. This geographical skew could hinder the generalization of our estimations for these longitudinal regions. \commentnj{We were able to highlight that thyroid cancer could potentially contribute to the elevated total cancer rate on the east coast by showing high incidence rates in this region. However, given the intricate nature of environmental factors influencing cancer development, coupled with numerous unmeasured potential exposures or risk factors, we are unable to definitively explain this phenomenon.} 

Another limitation pertains to the ecological nature of the data analyses. This approach does not allow for the quantification of risk factors, including position in the time zone, on the development of cancer using individual level data. \commentnj{Additionally, our analysis is hindered by missing reported data from certain counties, and a more limited range of cancer types compared to the study done by Gu et al. For instance, no reported cancer incidence rates are available for Kansas and Minnesota, and numerous counties lack data on specific cancer types. For example, we only have data on liver \& bile duct cancer incidence for 1117 counties. Addressing this issue would require access to more data, which represents a potential avenue for improvement in future studies.}

\commentnj{In conclusion, this study offers valuable insights into the association between cancer incidence rate and longitudinal positions. Our findings contribute to the ongoing discussion surrounding the implementation of daylight saving time. Further research is necessary to understand the underlying causes behind the observed variations in cancer rates across different regions within the United States. Such research can pave the way for developing targeted interventions that aim at preventing these cancers and improving public health.}

%% file: Sections/Acknowledgements.tex
Jin Niu and Charlotte Brown are supported in part by NSF Grant DMS-1646108.  

Michael Law is supported in part by NSF Grants DMS-1646108, DMS-2113364, and DMS-2203012.  

Justin Colacino is supported in part by NIH Grants R01-ES028802 and P30-CA046592.  

Ya\hspace{-.1em}'\hspace{-.1em}acov Ritov is supported in part by NSF Grant DMS-2113364.

%% file: Sections/Tables.tex
\begin{table}[htb] \centering 
  \caption{Reported Coefficients by Relative Position for All Cancer
            (with 95\% Confidence Interval)} 
  \label{linearmod_relpos} 
\begin{adjustbox}{width=0.8\textwidth}
{\begin{tabular}{@{\extracolsep{5pt}}lD{.}{.}{-3} D{.}{.}{-3} } 
\\[-1.8ex]\hline 
\hline \\[-1.8ex] 
 & \multicolumn{2}{c}{\textit{Dependent variable:}} \\ 
\cline{2-3} 
\\[-1.8ex] & \multicolumn{2}{c}{Cancer Incidence Rate} \\ 
 & \multicolumn{1}{c}{Coefficient} & \multicolumn{1}{c}{95\% CI} \\ 
\hline \\[-1.8ex] 
 Relative Position & -0.400 &(-0.898$, $0.098) \\ 
  Latitude & 1.541 & (0.616$, $2.465)^{***} \\ 
  High School & 156.030 & (91.022$, $221.039)^{***} \\ 
  Some College & 264.862 & (209.431$, $320.294)^{***} \\ 
  College and Above & 115.855 & (57.949$, $173.761)^{***} \\ 
  Elevation & -0.029 & (-0.035$, $-0.022)^{***} \\ 
  Medical Doctor pc & 1,970.561 & (1,176.039$, $2,765.083)^{***} \\ 
  Median Income & 0.0002 & (0.0001$, $0.0003)^{***} \\ 
  Obesity Rate & 54.728 & (17.158$, $92.299)^{***} \\ 
  Smoking Rate & 170.564 & (67.726$, $273.401)^{***} \\ 
  PM2.5 (air pollution) & 0.813 & (0.039$, $1.588)^{**} \\ 
  Water Violation & 2.308 & (-0.050$, $4.667)^{*} \\ 
  Race (White) & -403.091 & (-584.759$, $-221.424)^{***} \\ 
  Race (Black) & -398.755 & (-581.777$, $-215.733)^{***} \\ 
  Race (Native) & -552.755 & (-745.264$, $-360.245)^{***} \\ 
  Race (Asian) & -550.755 & (-744.413$, $-357.097)^{***} \\ 
  Race (Hispanic) & -447.579 & (-631.144$, $-264.015)^{***} \\ 
 \hline \\[-1.8ex] 
Observations & \multicolumn{2}{c}{2,853} \\ 
R$^{2}$ & \multicolumn{2}{c}{0.650} \\ 
Adjusted R$^{2}$ & \multicolumn{2}{c}{0.642} \\ 
Residual Std. Error (df = 2791) & \multicolumn{2}{c}{0.774} \\ 
F Statistic (df = 61; 2791) & \multicolumn{2}{c}{84.954$^{***}$} \\ 
\hline 
\hline \\[-1.8ex] 
\textit{Note:}  & \multicolumn{2}{r}{$^{*}$p$<$0.1; $^{**}$p$<$0.05; $^{***}$p$<$0.01} \\ 
\end{tabular}}
\end{adjustbox}
\end{table} 


%% file: Sections/Figure_Legend.tex
\setcounter{table}{0}

\begin{table}[!htb] \centering
    \caption{Figure Legend for Figures in Main Text} 
\begin{adjustbox}{width = \textwidth}
{\begin{tabular}{|c|c|c|c|}
\hline
Fig no. & Title                                                                                                              & Summary                                                                                                                                                                                                                                   & Subfigure Description                                                                                                                                                                                                                                                                                       \\ \hline
Fig. 1  & \begin{tabular}[c]{@{}c@{}}Map of Relative Position \\ and Cancer Incidence\end{tabular}                           & \begin{tabular}[c]{@{}c@{}}Figure 1 shows maps of relative \\ position within time zone and \\ cancer incidence rate by county \\ for 607 and 2853 counties \\ (observations), respectively.\end{tabular}                                 & \begin{tabular}[c]{@{}c@{}}(a) Map of Relative Position \\ within Time Zone for 607 counties;\\ (b) Map of Relative Position \\ within Time Zone for 2853 counties;\\ (c) Map of U.S. Cancer Incidence \\ Rate for 607 counties;\\ (d) Map of U.S. Cancer Incidence \\ Rate for 2853 counties;\end{tabular} \\ \hline
Fig. 2  & \begin{tabular}[c]{@{}c@{}}Output of Linear \\ Approximation \\ for Total Cancers\end{tabular}                 & \begin{tabular}[c]{@{}c@{}}Figure 2 shows linear approximation \\ results of total cancer incidence \\ rate by relative position with \\ 95\% bootstrap confidence band.\end{tabular}                                                 & \begin{tabular}[c]{@{}c@{}}(a) with 607 counties, \\ (b) with 2853 counties\end{tabular}                                                                                                                                                                                                                    \\ \hline
Fig. 3  & \begin{tabular}[c]{@{}c@{}}Output of Natural Splines \\ for Hormonally Associated \\ Cancer Incidence\end{tabular} & \begin{tabular}[c]{@{}c@{}}Figure 3 shows the output \\ of natural splines conducted on \\ incidence by longitude \\ and time zone for four of the \\ hormonally associated cancers, with \\ 95\% bootstrap confidence band.\end{tabular} & \begin{tabular}[c]{@{}c@{}}(a)  Breast Cancer (n = 2615),\\ (b)  Ovary Cancer (n = 890),\\ (c)  Prostate Cancer (n = 2623),\\ (d)  Thyroid Cancer (n = 1191)\end{tabular}                                                                                                                                   \\ \hline
Fig. 4  & \begin{tabular}[c]{@{}c@{}}Output of Natural Splines \\ for Most Prevalent Cancers\end{tabular}                    & \begin{tabular}[c]{@{}c@{}}Figure 4 shows the output \\ of natural splines conducted on \\ incidence by longitude \\ and time zone for four of the \\ most prevalent cancers, with \\ 95\% bootstrap confidence band.\end{tabular}        & \begin{tabular}[c]{@{}c@{}}(a)  Liver \& Bile Duct Cancer \\ (without MST, n = 1063),\\ (b)  Lung \& Bronchus Cancer (n = 2441),\\ (c) Colon \& Rectum Cancer (n = 1955)\\ (d)  Pancreas Cancer (n = 1191)\end{tabular}                                                                                     \\ \hline
Fig. 5  & \begin{tabular}[c]{@{}c@{}}Output of Natural Splines \\ for Total Cancers\end{tabular}                         & \begin{tabular}[c]{@{}c@{}}Figure 5 shows the output \\ of natural splines conducted on \\ incidence by longitude and time zone \\ for total cancer, with 95\% \\ Bootstrap Confidence Band (n = 2853).\end{tabular}                  & \begin{tabular}[c]{@{}c@{}}(a)  Point-wise CI,\\ (b)  Uniform CI Restricted to \\ Counties in Overlapping Regions\end{tabular}                                                                                                                                                                              \\ \hline
\end{tabular}}
\end{adjustbox}
\end{table}

%% file: Sections/Figures.tex
\begin{figure}[htb]
    \centering
    \begin{subfigure}[tb]{0.45\textwidth}
        \centering
        \includegraphics[width=\textwidth]{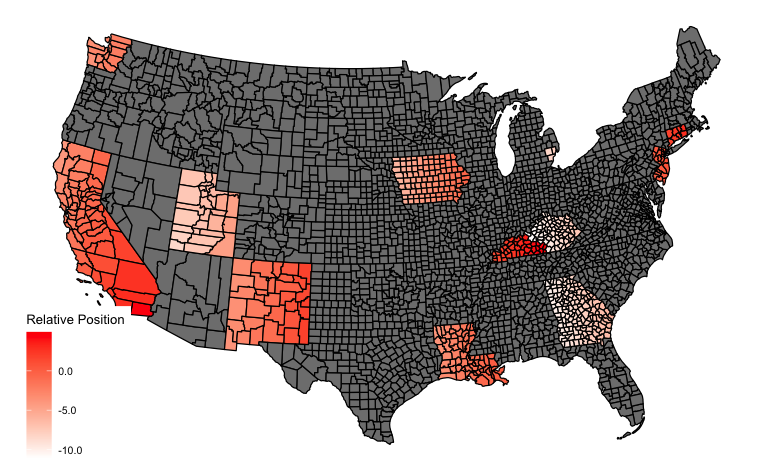}
        \caption{}
        \label{fig:rel_position}
    \end{subfigure}
    \hfill
    \begin{subfigure}[tb]{0.45\textwidth}
        \centering
        \includegraphics[width=\textwidth]{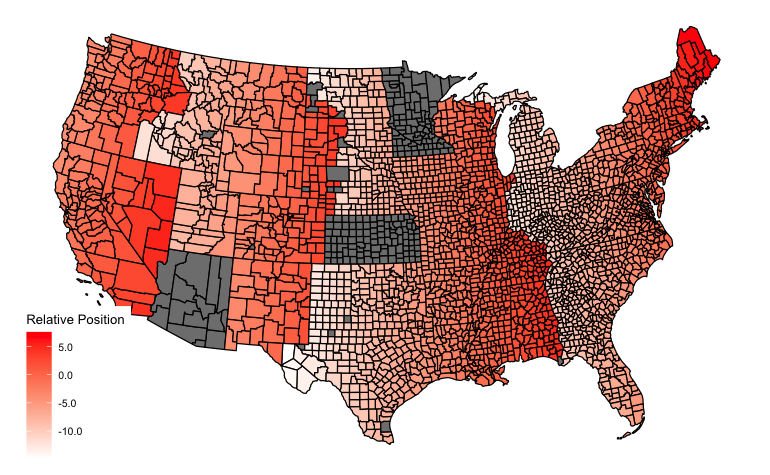}
        \caption{}
        \label{fig:rel_position}
    \end{subfigure}
    \\
    \begin{subfigure}[tb]{0.45\textwidth}
        \centering
        \includegraphics[width=\textwidth]{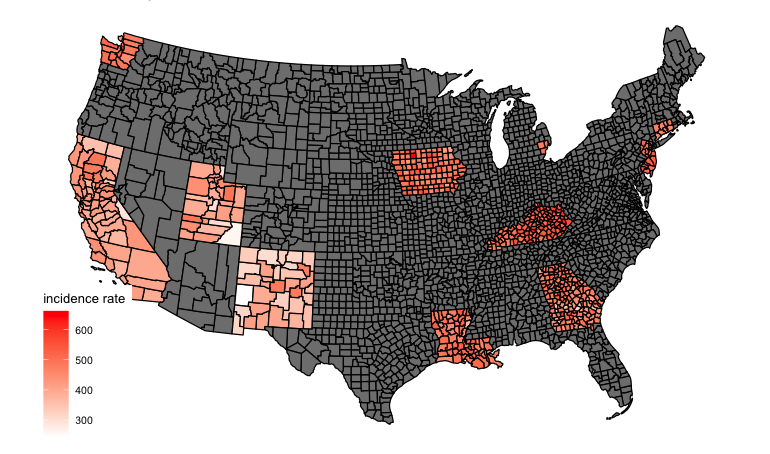}
        \caption{}
        \label{fig:cancerrate_map}
    \end{subfigure}
    \hfill
    \begin{subfigure}[tb]{0.45\textwidth}
        \centering
        \includegraphics[width=\textwidth]{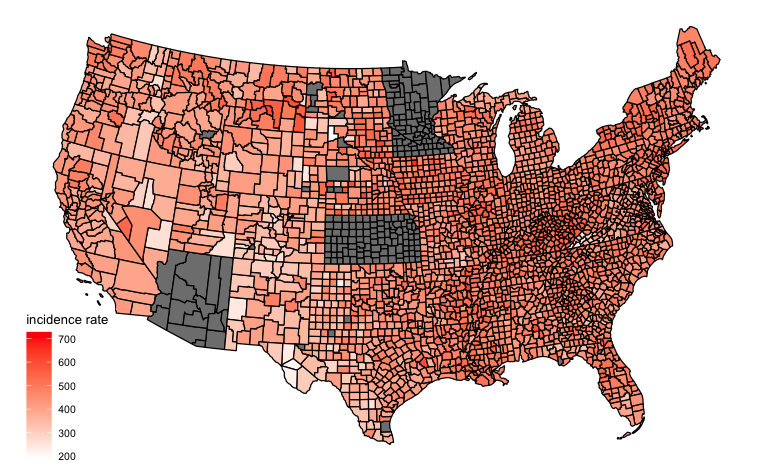}
        \caption{}
        \label{fig:cancerrate_map}
    \end{subfigure}
    \caption{}
    \label{fig:usmap_relpos}
\end{figure}

\clearpage
\begin{figure}[htb]
    \centering
    \begin{subfigure}[tb]{0.45\textwidth}
         \centering
         \includegraphics[width=\textwidth]{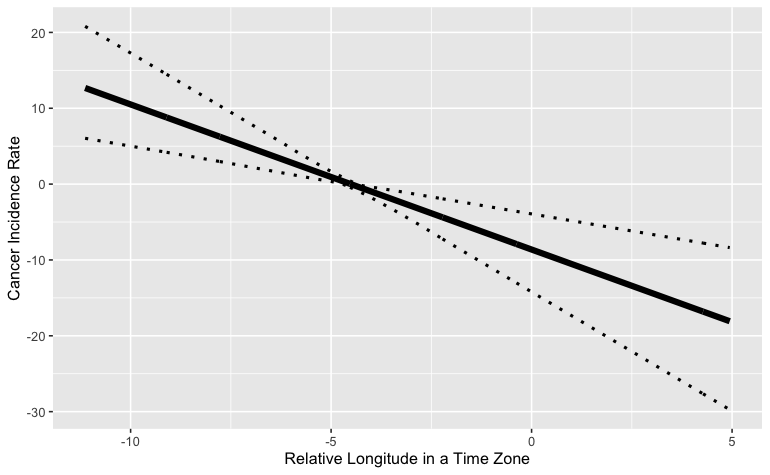}
         \caption{}
         \label{fig:607linear}
    \end{subfigure}
    \hfill
    \begin{subfigure}[tb]{0.45\textwidth}
         \centering
         \includegraphics[width=\textwidth]{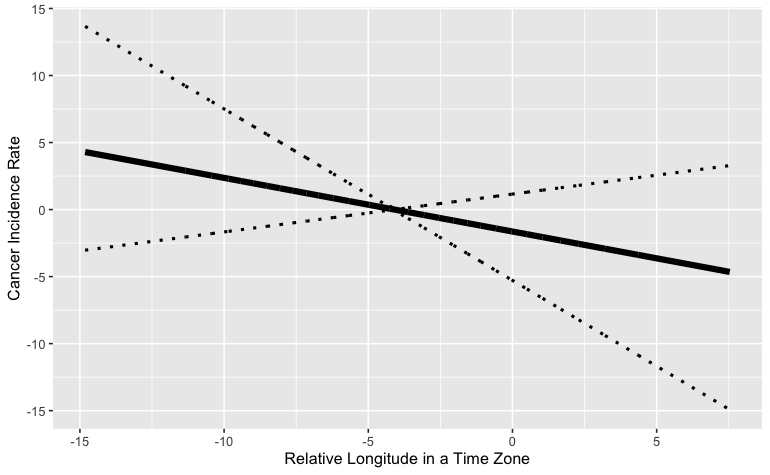}
         \caption{}
         \label{fig:alllinear}
    \end{subfigure}
    \caption{}
    \label{fig:linear_approx}
\end{figure}

\clearpage
\begin{figure}[htb]
    \centering
    \begin{subfigure}[tb]{0.45\textwidth}
        \centering
        \includegraphics[width = \textwidth]{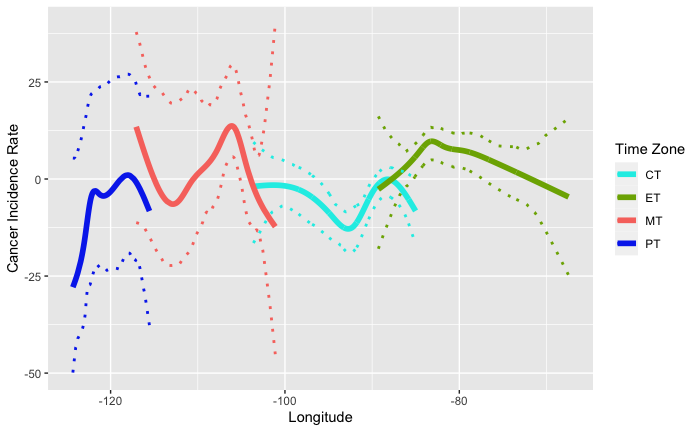}
        \caption{}
        \label{br_spl}
    \end{subfigure}
    \hfill
    \begin{subfigure}[tb]{0.45\textwidth}
        \centering
        \includegraphics[width = \textwidth]{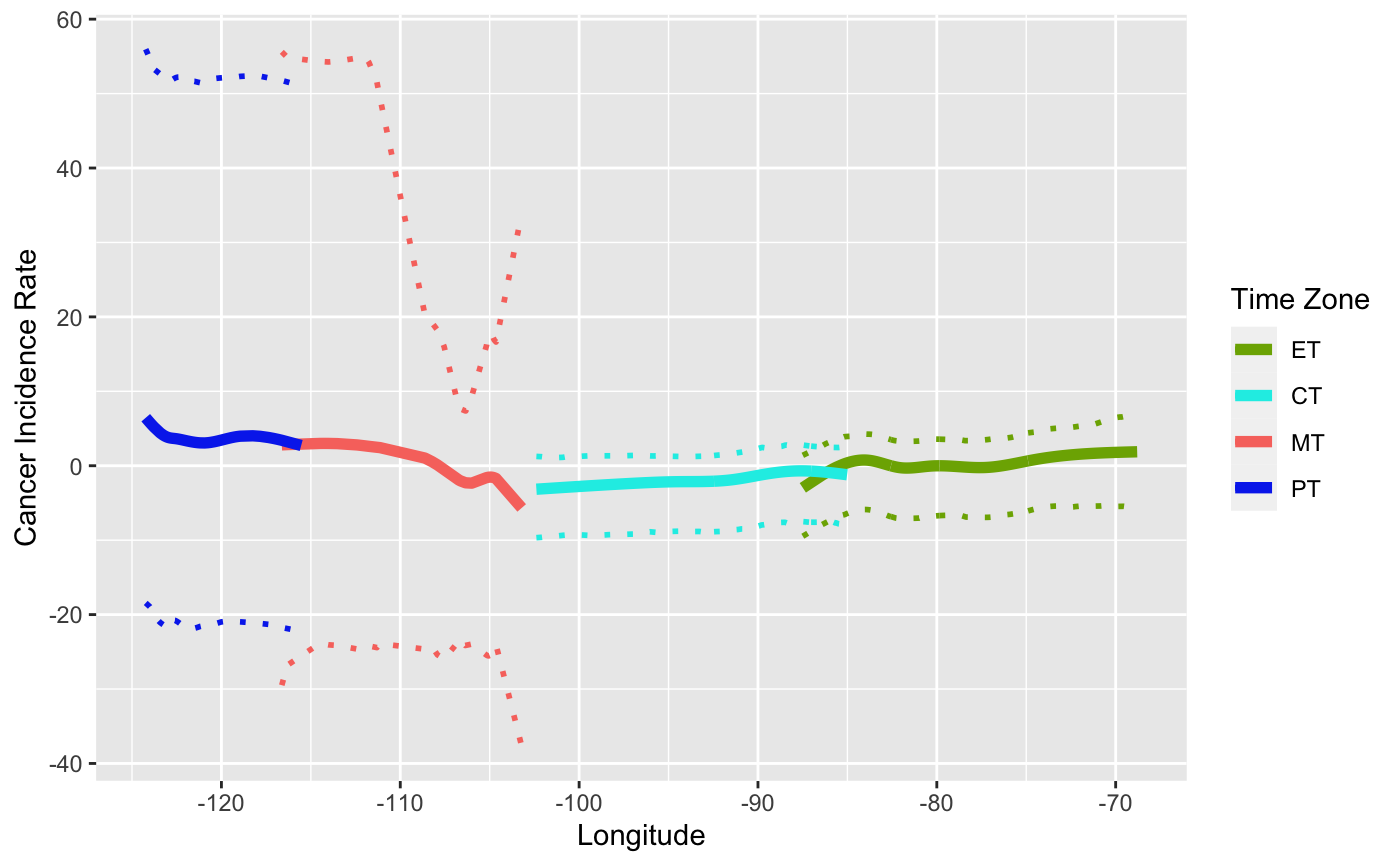}
        \caption{}
        \label{ova_spl}
    \end{subfigure} \\
    \begin{subfigure}[tb]{0.45\textwidth}
        \centering
        \includegraphics[width = \textwidth]{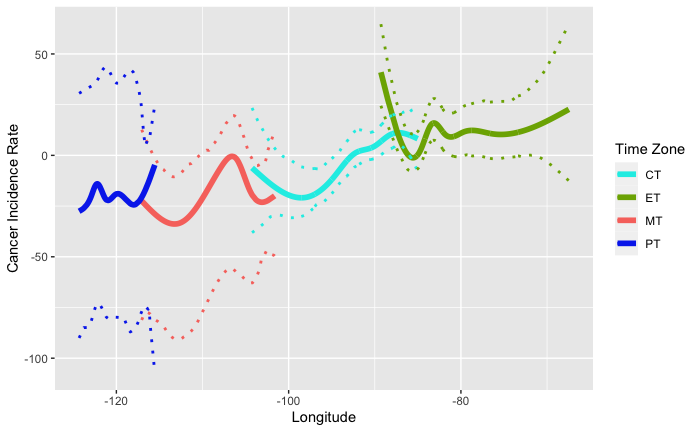}
        \caption{}
        \label{ps_spl}
    \end{subfigure}
    \hfill
    \begin{subfigure}[tb]{0.45\textwidth}
        \centering
        \includegraphics[width = \textwidth]{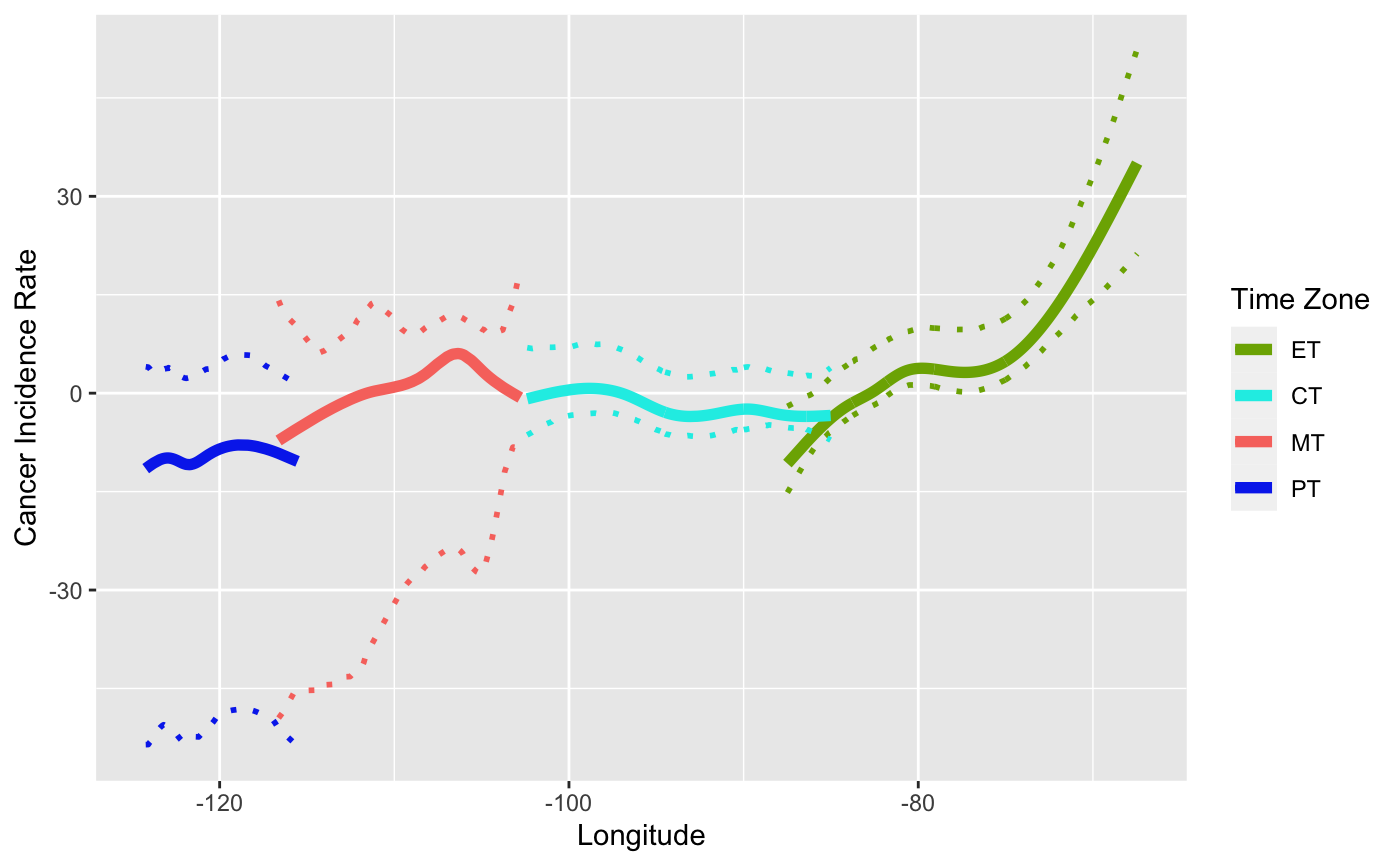}
        \caption{}
        \label{thy_spl}
    \end{subfigure}
    \caption{}
    \label{fig:hormon_boot}
\end{figure}

\clearpage
\begin{figure}[htb]
    \centering
    \begin{subfigure}[tb]{0.45\textwidth}
        \centering
        \includegraphics[width = \textwidth]{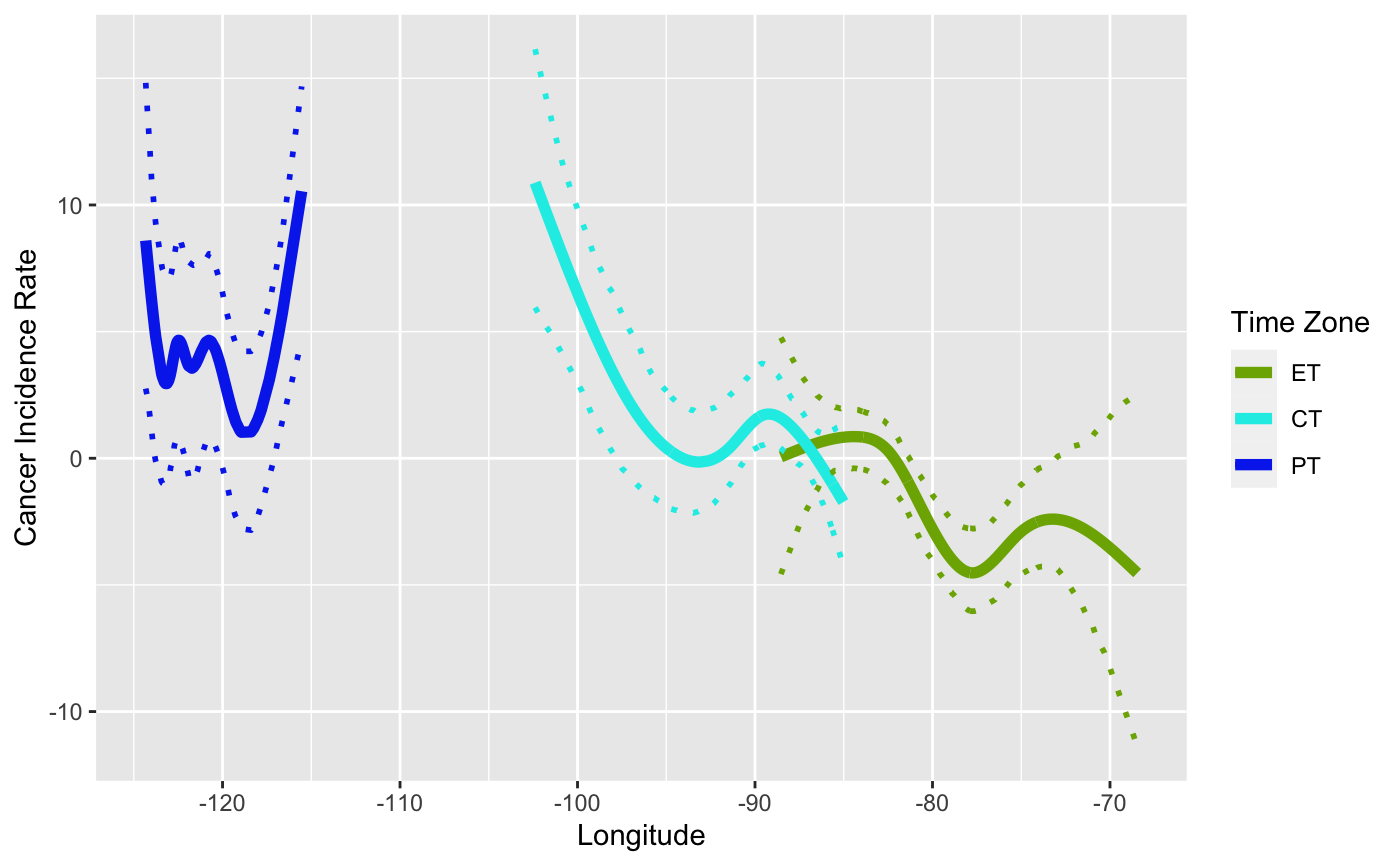}
        \caption{}
        \label{liver_spl}
    \end{subfigure}
    \hfill
    \begin{subfigure}[tb]{0.45\textwidth}
        \centering
        \includegraphics[width = \textwidth]{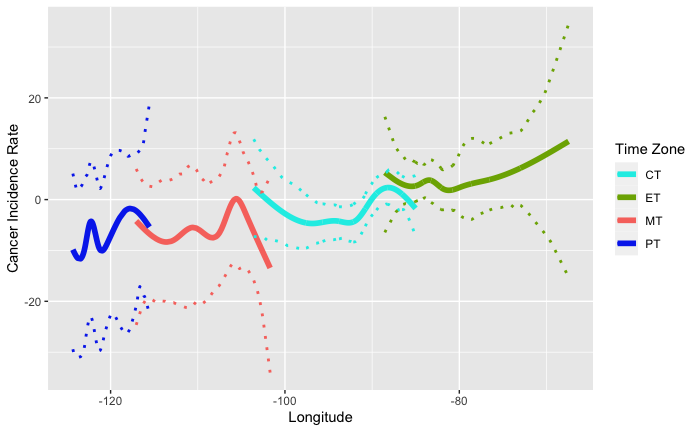}
        \caption{}
        \label{lung_spl}
    \end{subfigure} \\
    \begin{subfigure}[tb]{0.45\textwidth}
        \centering
        \includegraphics[width = \textwidth]{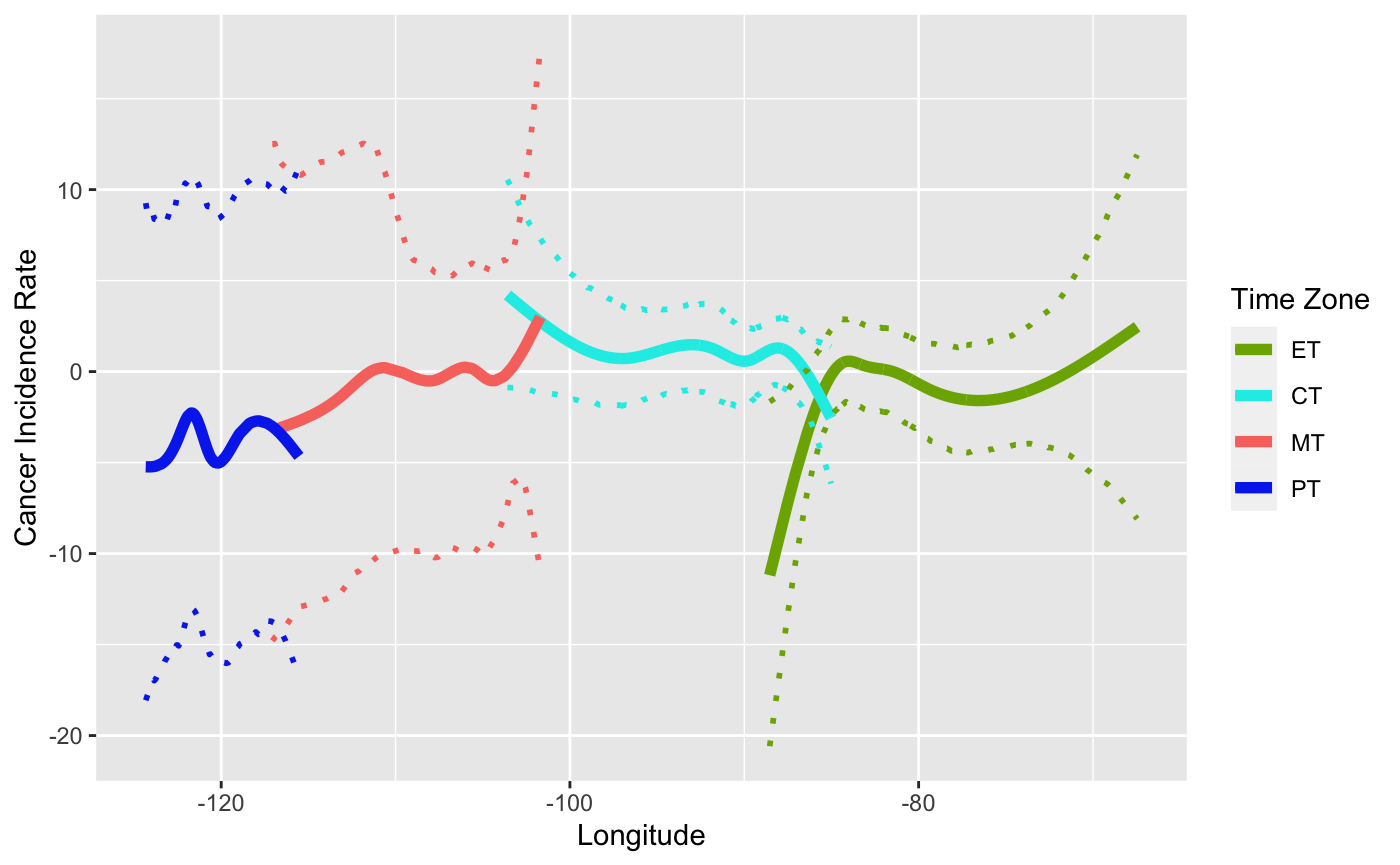}
        \caption{}
        \label{col_spl}
    \end{subfigure}
    \hfill
    \begin{subfigure}[tb]{0.45\textwidth}
        \centering
        \includegraphics[width = \textwidth]{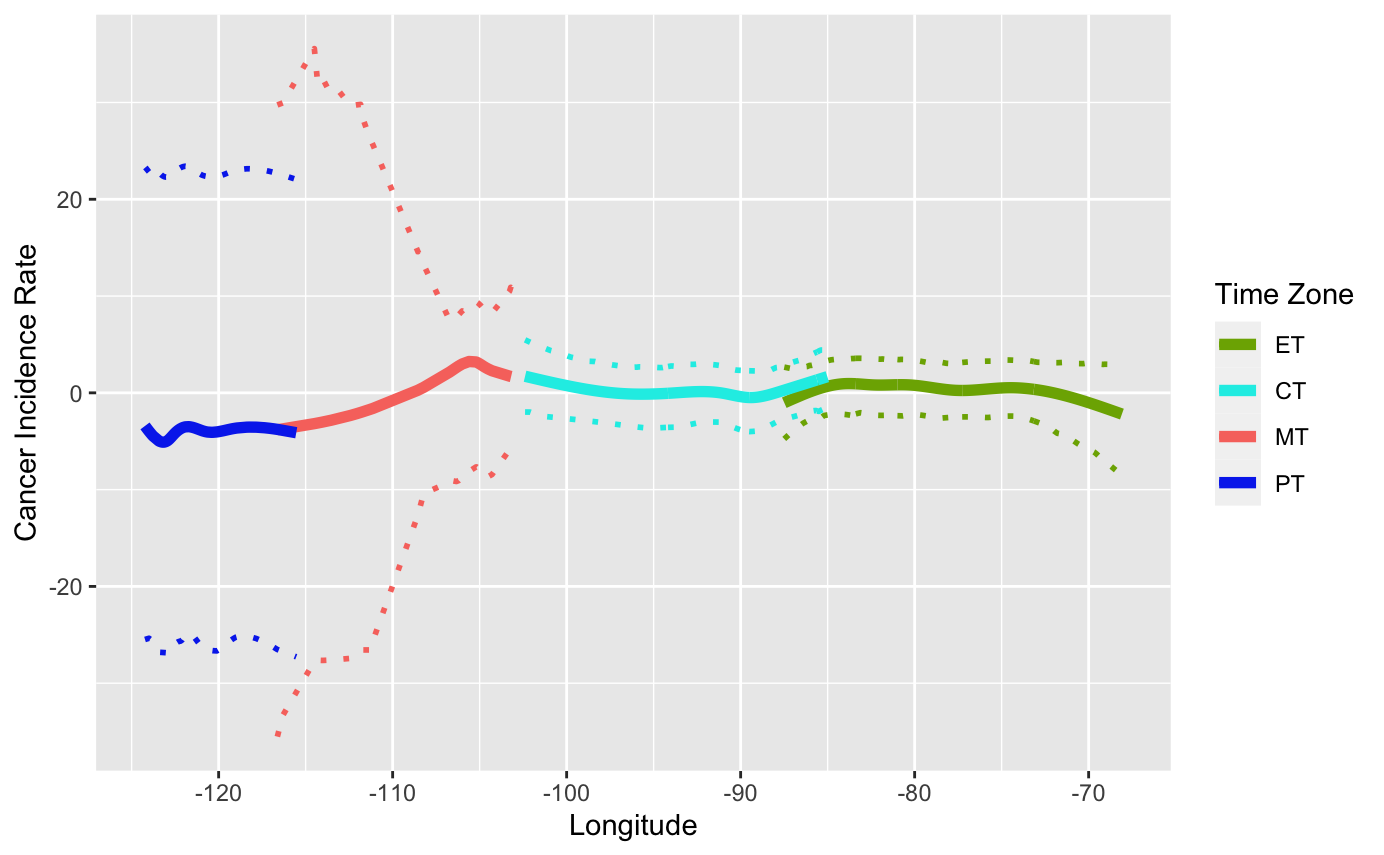}
        \caption{}
        \label{pan_spl}
    \end{subfigure}
    \caption{}
    \label{fig:subcancer_boot}
\end{figure}

\clearpage
\begin{figure}[htb]
    \centering
    \begin{subfigure}[tb]{0.7\textwidth}
        \centering
        \includegraphics[width=\textwidth]{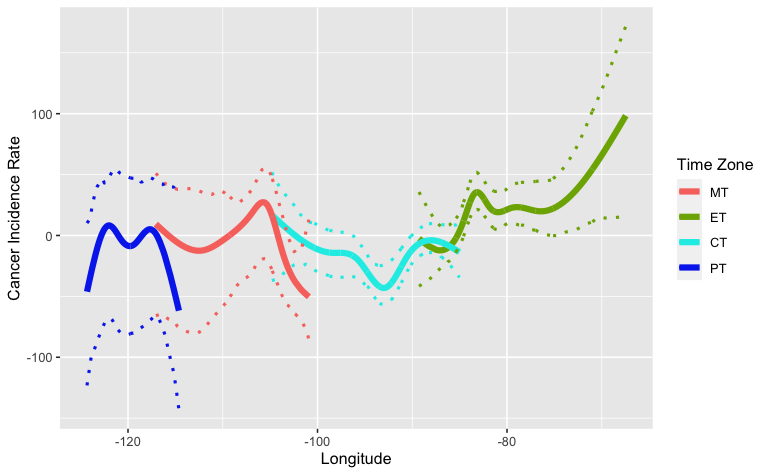}
        \caption{}
        \label{fig:comp_spl_ptwise}
    \end{subfigure}
    \\
    \begin{subfigure}[tb]{0.7\textwidth}
        \centering
        \includegraphics[width=\textwidth]{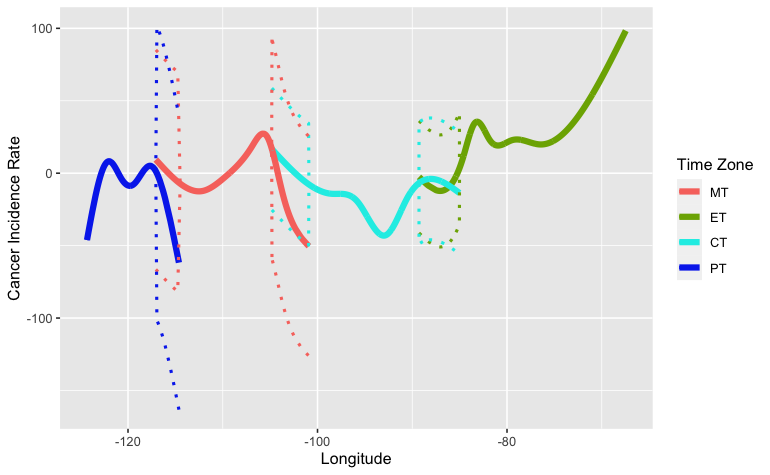}
        \caption{}
        \label{fig:comp_spl_unif}
    \end{subfigure}    
    \caption{}
    \label{fig:comp_spl}
\end{figure}

%% file: Sections/supplementary.tex
\begin{figure}[htb]
    \centering
    \includegraphics[width=5in]{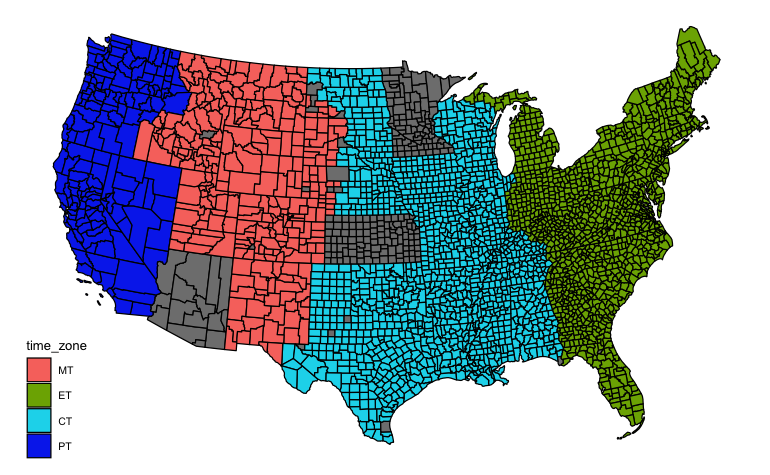} \\
    \caption{U.S. Counties By Time Zone ($n=2853$)}{Supplementary Figure 1 shows a map of U.S. counties contained in our analysis of total cancer incidence according to their respective time zones ($n = 2853$).}
    \label{fig:timezone_map}
\end{figure}

\clearpage
\begin{figure}[htb]
    \centering
    \begin{subfigure}[tb]{0.45\textwidth}
        \centering
        \includegraphics[width = \textwidth]{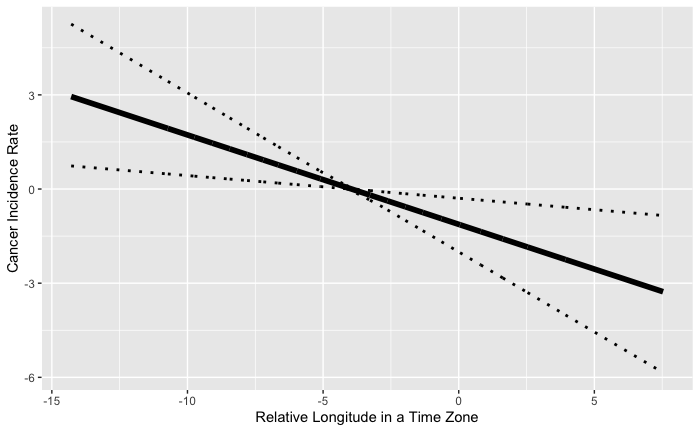}
        \caption{Breast Cancer ($n = 2615$)}
        \label{}
    \end{subfigure}
    \hfill
    \begin{subfigure}[tb]{0.45\textwidth}
        \centering
        \includegraphics[width = \textwidth]{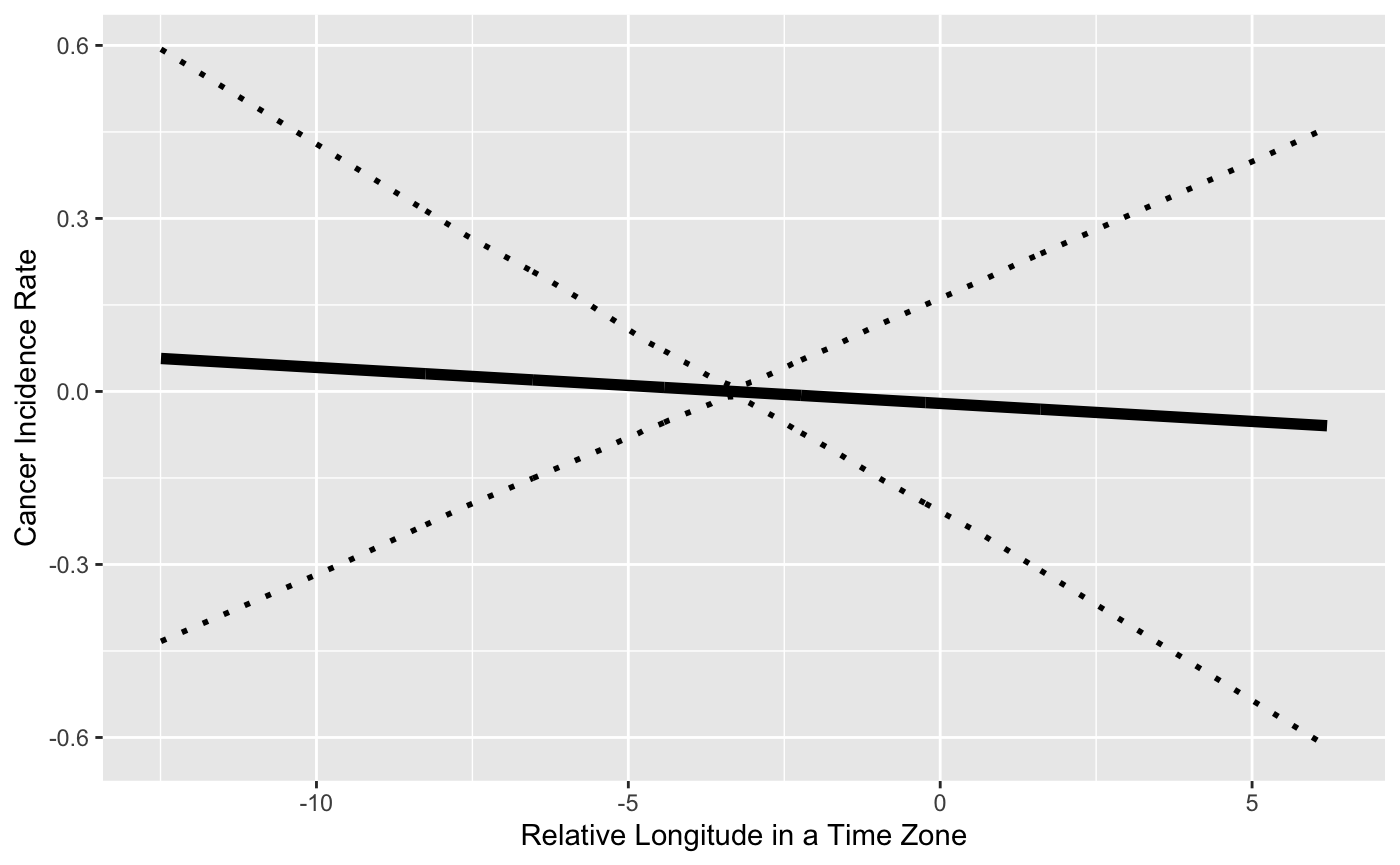}
        \caption{Ovary Cancer ($n = 890$)}
        \label{}
    \end{subfigure} \\
    \begin{subfigure}[tb]{0.45\textwidth}
        \centering
        \includegraphics[width = \textwidth]{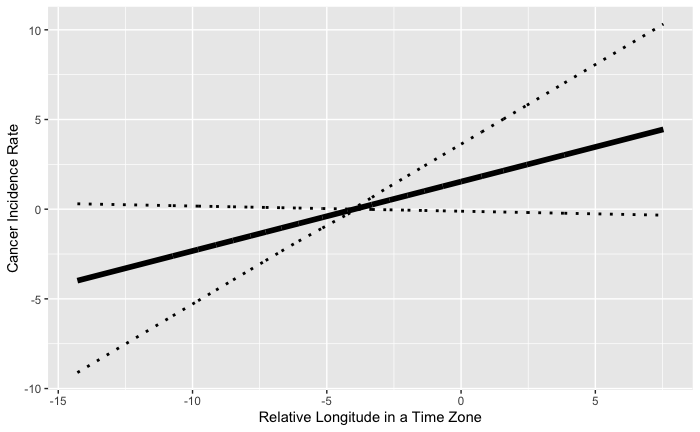}
        \caption{Prostate Cancer ($n = 2623$)}
        \label{ps_lin}
    \end{subfigure}
    \hfill
    \begin{subfigure}[tb]{0.45\textwidth}
        \centering
        \includegraphics[width = \textwidth]{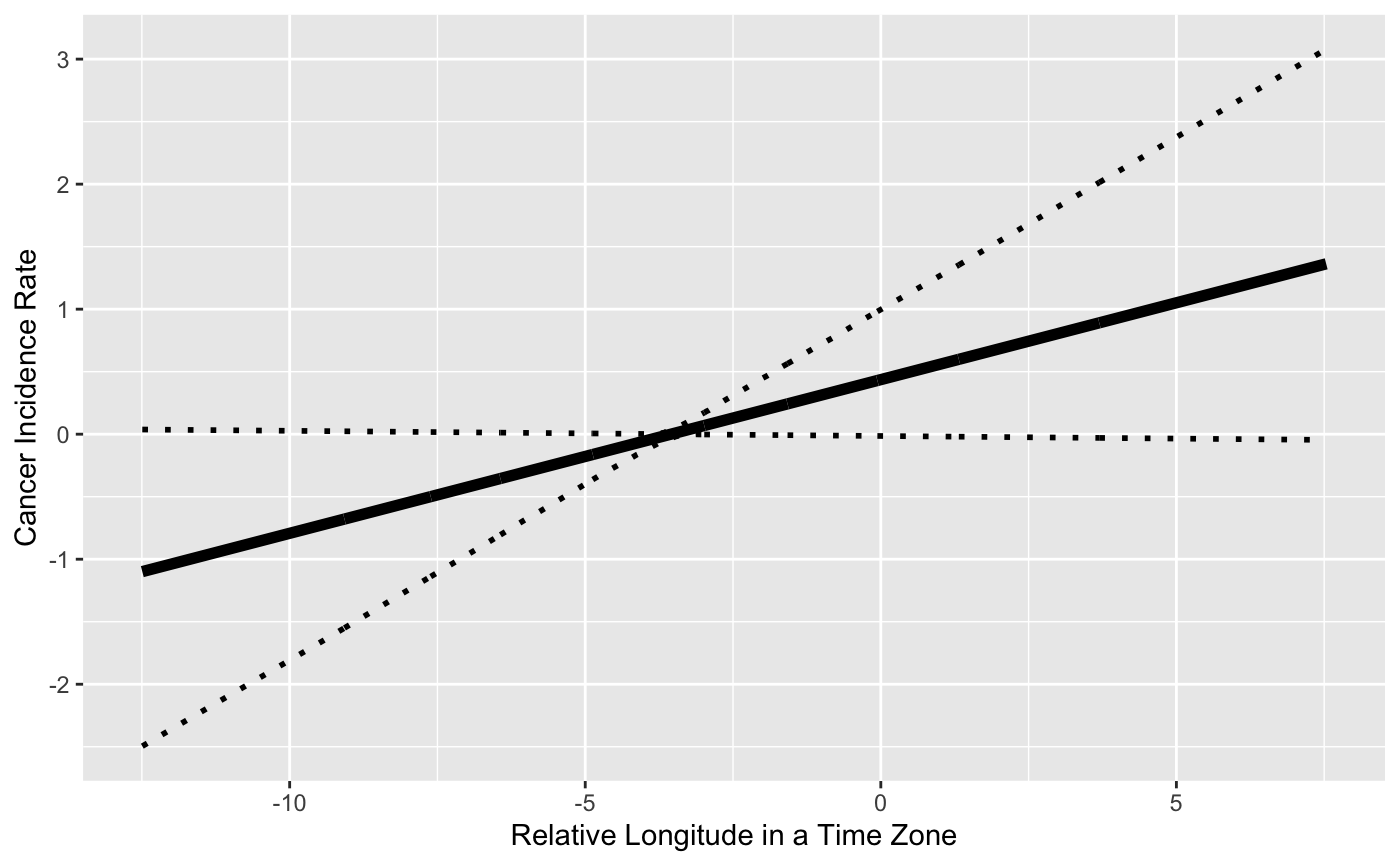}
        \caption{Thyroid Cancer ($n = 1191$)}
        \label{thy_lin}
    \end{subfigure}
    \caption{Output of Linear Approximation for Hormonally Associated Cancer Incidence}{Supplementary Figure 2 shows the linear approximation result of incidence by relative position for four of the hormonally associated cancers, with 95\% bootstrap confidence band.}
    \label{fig:hormon_lin}
\end{figure}

\clearpage
\begin{figure}[htb]
    \centering
    \begin{subfigure}[tb]{0.45\textwidth}
        \centering
        \includegraphics[width = \textwidth]{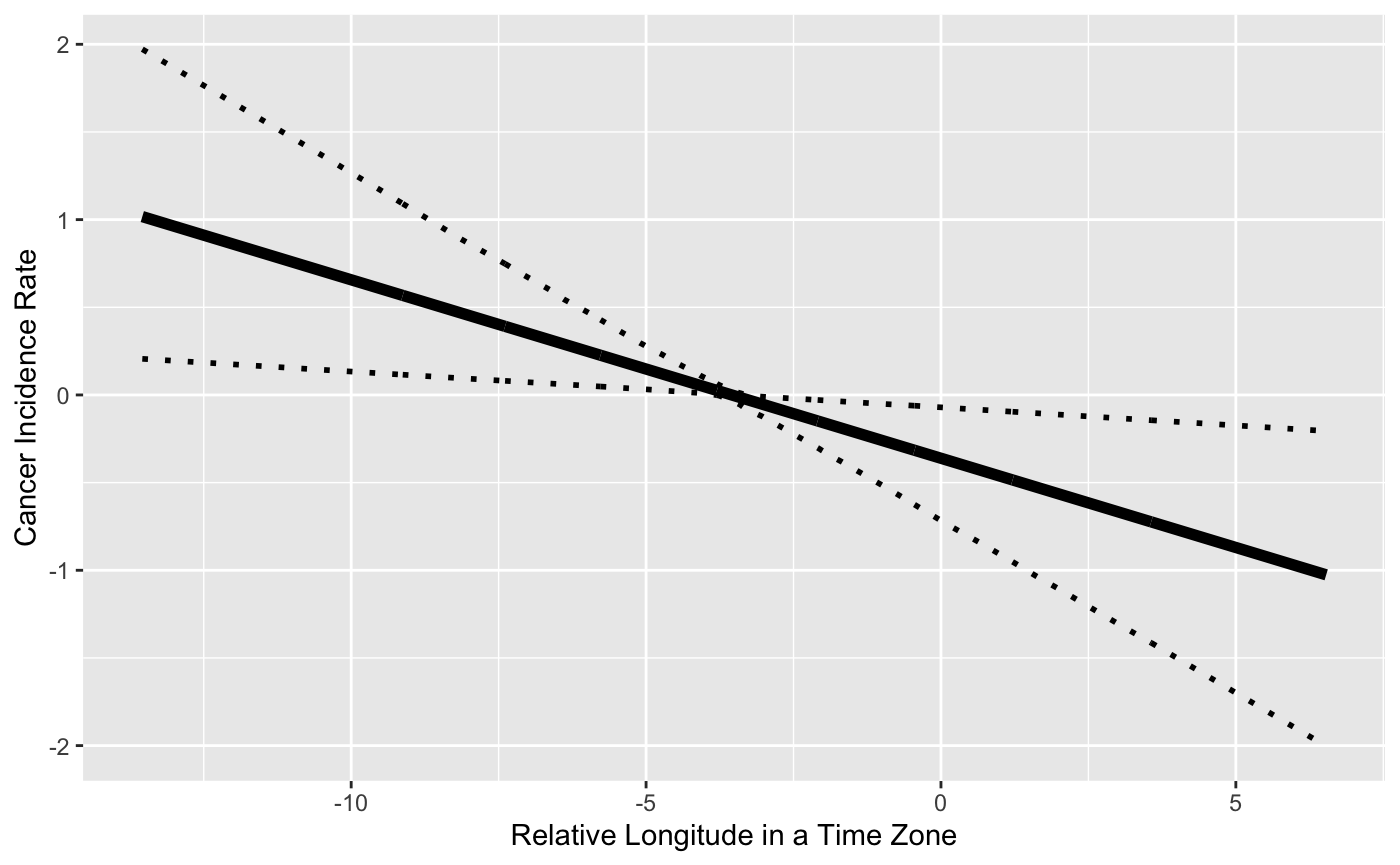}
        \caption{Liver \& Bile Duct Cancer (without MST, $n = 1063$)}
        \label{liver_lin_noMT}
    \end{subfigure}
    \hfill
    \begin{subfigure}[tb]{0.45\textwidth}
        \centering
        \includegraphics[width = \textwidth]{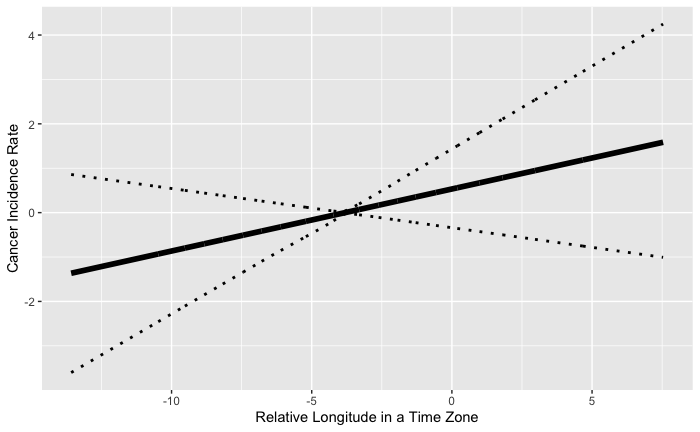}
        \caption{Lung \& Bronchus Cancer ($n = 2441$)}
        \label{lung_lin}
    \end{subfigure} \\
    \begin{subfigure}[tb]{0.45\textwidth}
        \centering
        \includegraphics[width = \textwidth]{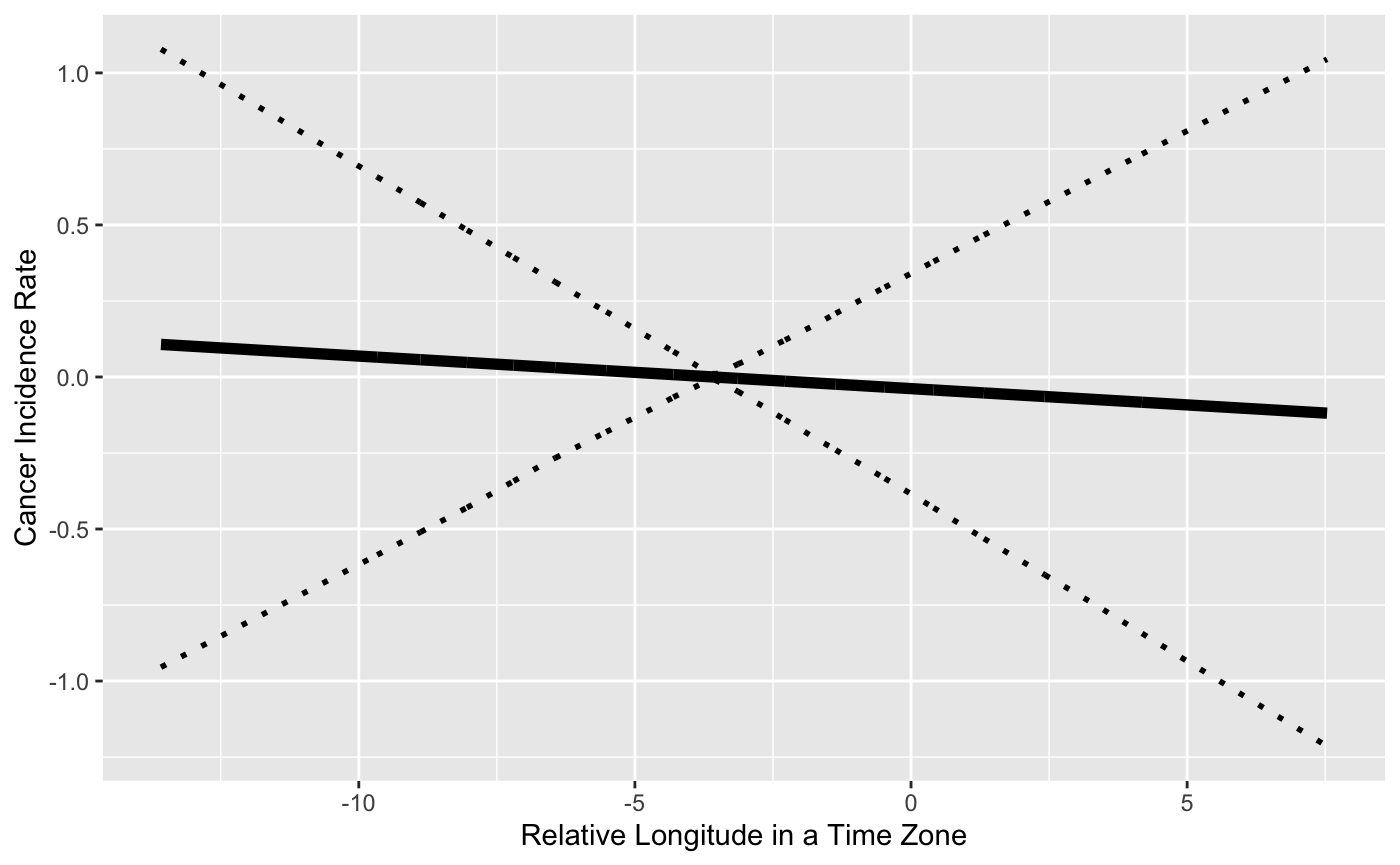}
        \caption{Colon \& Rectum Cancer ($n = 1955$)}
        \label{col_lin}
    \end{subfigure}
    \hfill
    \begin{subfigure}[tb]{0.45\textwidth}
        \centering
        \includegraphics[width = \textwidth]{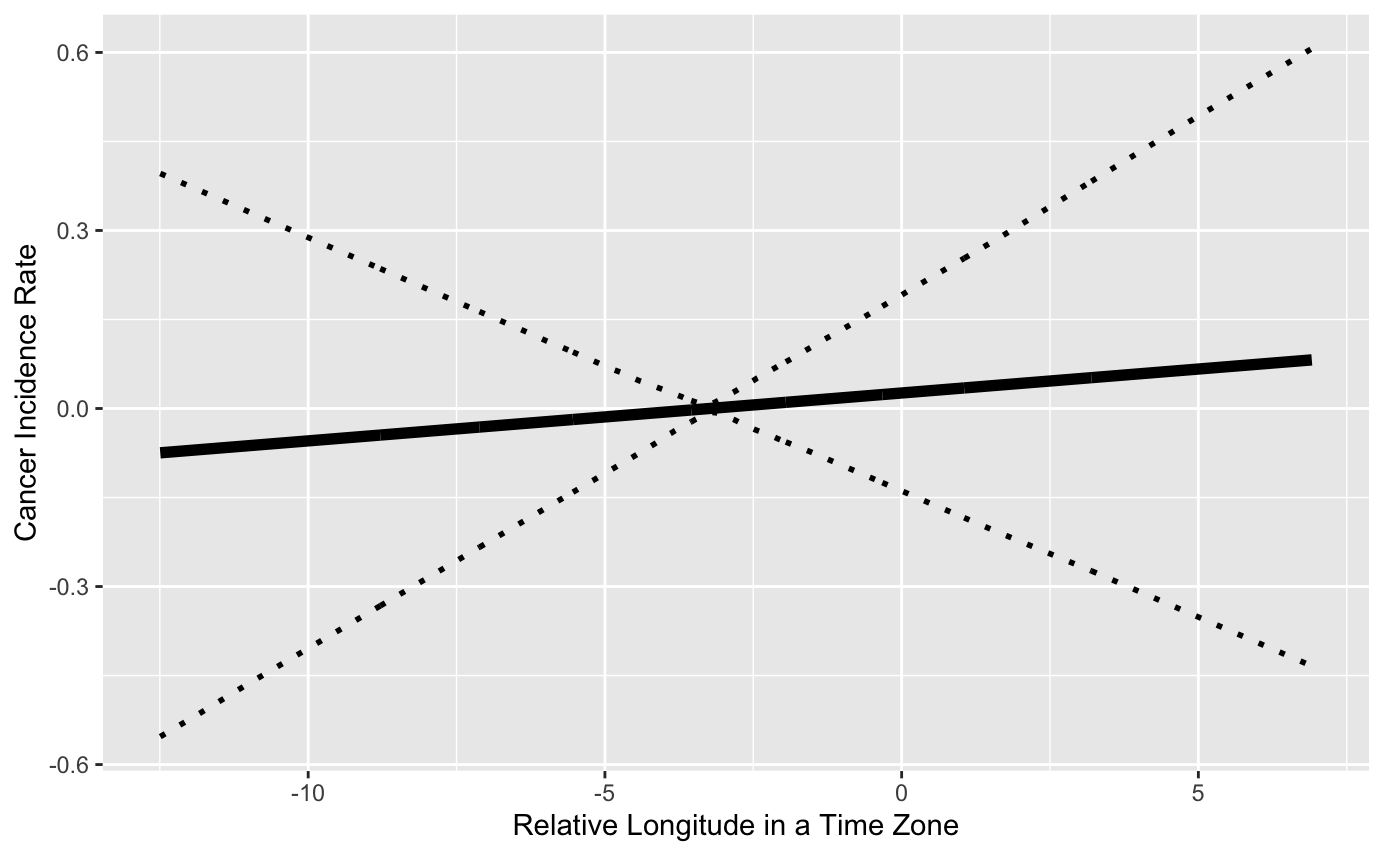}
        \caption{Pancreas Cancer ($n = 1191$)}
        \label{pan_lin}
    \end{subfigure}
    \caption{Output of Linear Approximation for Most Prevalent Cancer Incidence}{Supplementary Figure 3 shows the linear approximation result of incidence by relative position for four of the most prevalent cancers, with 95\% bootstrap confidence band.}
    \label{fig:subcancer_lin}
\end{figure}

\clearpage
\begin{table}[htb] \centering 
  \caption{Summary Statistics for Selected Variables} 
  \label{sumstats} 
\begin{tabular}{@{\extracolsep{5pt}}lcccc} 
\\[-1.8ex]\hline 
\hline \\[-1.8ex] 
Statistic & \multicolumn{1}{c}{Mean} & \multicolumn{1}{c}{St. Dev.} & \multicolumn{1}{c}{Min} & \multicolumn{1}{c}{Max} \\ 
\hline \\[-1.8ex] 
Cancer Incidence Rate (total) & 452.523 & 57.648 & 211.200 & 661.300 \\ 
Relative Position & $-$4.091 & 4.557 & $-$14.820 & 7.534 \\ 
Below High School & 0.133 & 0.062 & 0.011 & 0.467 \\ 
High School & 0.344 & 0.073 & 0.078 & 0.574 \\ 
Some College & 0.305 & 0.050 & 0.112 & 0.473 \\ 
College and Above & 0.218 & 0.096 & 0.032 & 0.753 \\
Elevation (meters) & 393.404 & 465.509 & $-$21.000 & 3,163.000 \\ 
Medical Doctor per capita & 0.001 & 0.002 & 0.000 & 0.038 \\ 
Median Income & 48,579.740 & 13,163.530 & 21,087 & 220,645 \\ 
Obesity Rate & 33.722 & 5.876 & 11.000 & 58.900 \\ 
Smoking Rate & 21.340 & 4.240 & 7.076 & 40.937 \\ 
PM2.5 (air pollution) & 7.789 & 1.647 & 1.500 & 16.000 \\ 
Water Violation & 0.381 & 0.486 & 0 & 1 \\ 
Race (White) & 0.758 & 0.200 & 0.027 & 0.978 \\ 
Race (Black) & 0.095 & 0.146 & 0.000 & 0.859 \\ 
Race (Native) & 0.016 & 0.061 & 0.0004 & 0.892 \\ 
Race (Asian) & 0.014 & 0.025 & 0.000 & 0.384 \\ 
Race (Hispanic) & 0.098 & 0.140 & 0.006 & 0.964 \\ 
Race (Other) & 0.018 & 0.011 & 0.000 & 0.099 \\ 
\hline \\[-1.8ex] 
\multicolumn{5}{l}{\textit{Number of Observations:} 2853} \\ 
\end{tabular} 
\end{table} 

\clearpage
\begin{table}[htb] \centering
    \caption{Data Sources}
    \label{datasources}
\begin{tabular}{|c|c|}
\hline
Data    & Source   \\ \hline
Cancer incidence rate                                                                                     & \begin{tabular}[c]{@{}c@{}}National Cancer Institute and \\ Centers for Disease Control and Prevention\end{tabular}                \\ \hline
Educational Attainment                                                                                    & Economic Research Service, U.S. Department of Agriculture                                                                          \\ \hline
Race                                                                                                      & U.S. Census Bureau                                                                                                                 \\ \hline
Elevation                                                                                                 & Open Elevation                                                                             \\ \hline
Median income per capita                                                                                  & Bureau of Economic Analysis                                                                                                        \\ \hline
Population Center                                                                                         & U.S. Census Bureau                                                                                                                 \\ \hline
Access to healthcare                                                                                      & Health Resources \& Services Administration                                                                                        \\ \hline
\begin{tabular}[c]{@{}c@{}}Smoking rate \& obesity rate, \\ Air pollution \& water violation\end{tabular} & \begin{tabular}[c]{@{}c@{}}County Health Rankings \& Roadmaps, \\ University of Wisconsin Population Health Institute\end{tabular} \\ \hline
\end{tabular}
\end{table}

\clearpage
\begin{figure}[htb]
    \centering
    \begin{subfigure}[tb]{0.45\textwidth}
        \centering
        \includegraphics[width = \textwidth]{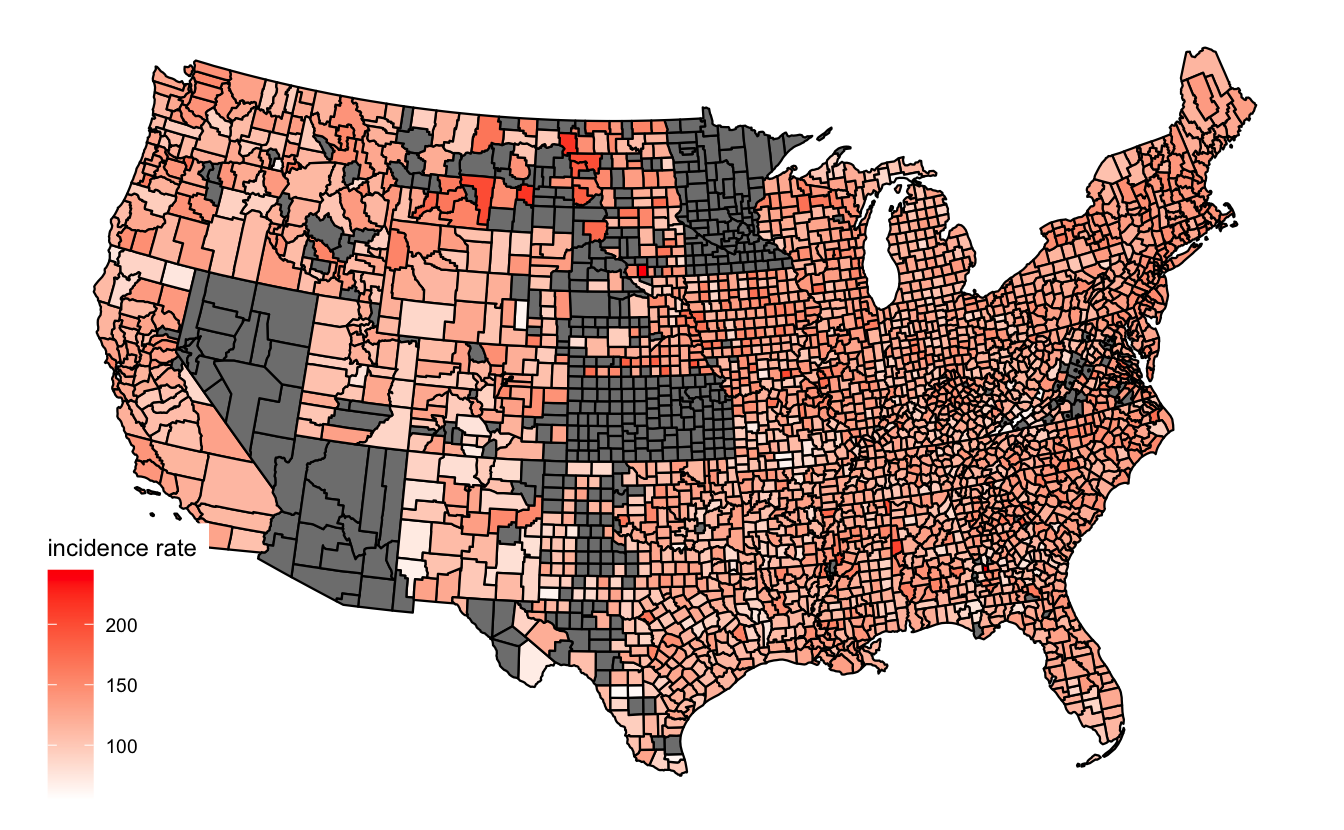}
        \caption{Breast Cancer ($n = 2615$)} 
        \label{}
    \end{subfigure}
    \hfill
    \begin{subfigure}[tb]{0.45\textwidth}
        \centering
        \includegraphics[width = \textwidth]{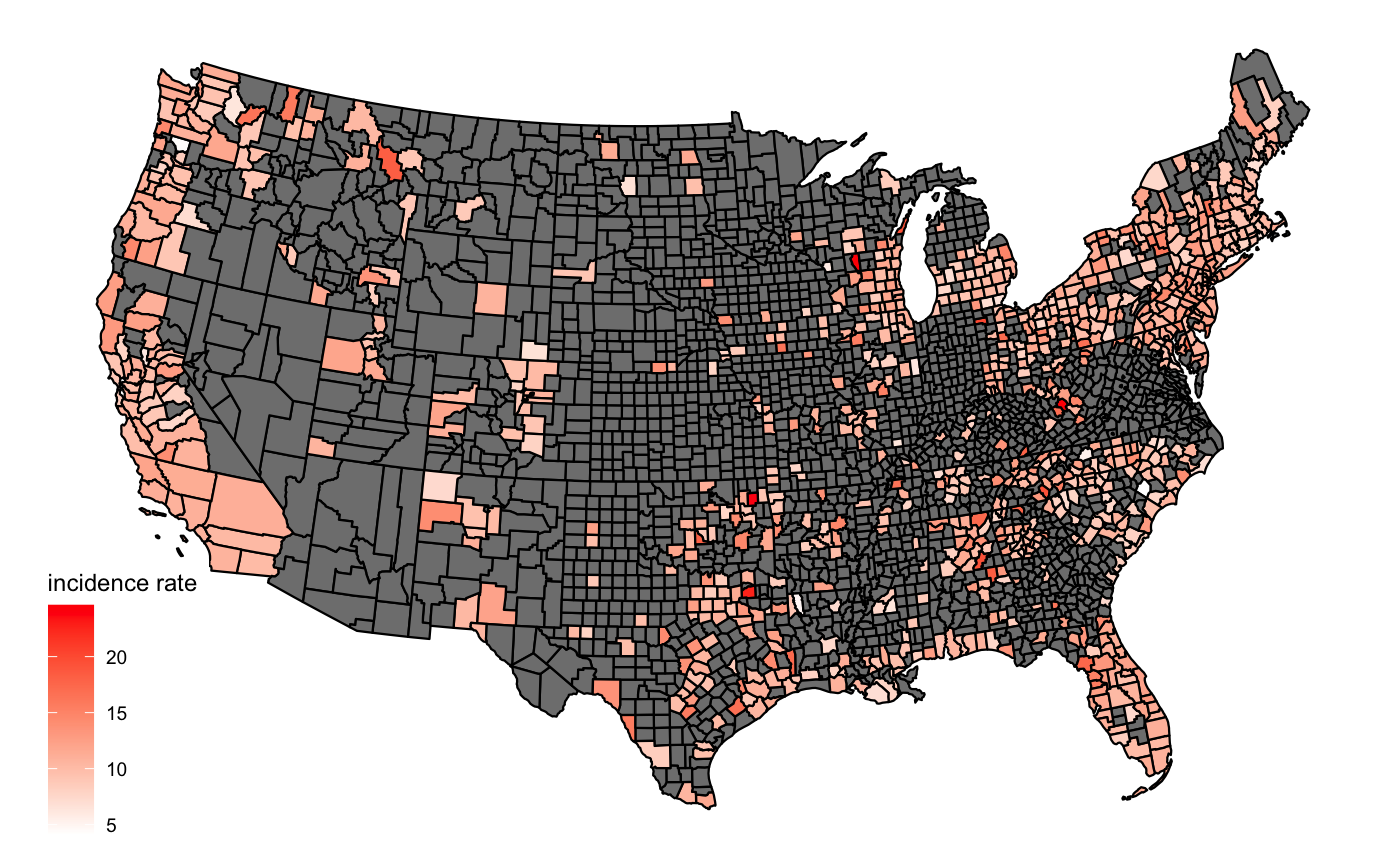}
        \caption{Ovary Cancer ($n = 890$)} 
        \label{}
    \end{subfigure} \\
    \begin{subfigure}[tb]{0.45\textwidth}
        \centering
        \includegraphics[width = \textwidth]{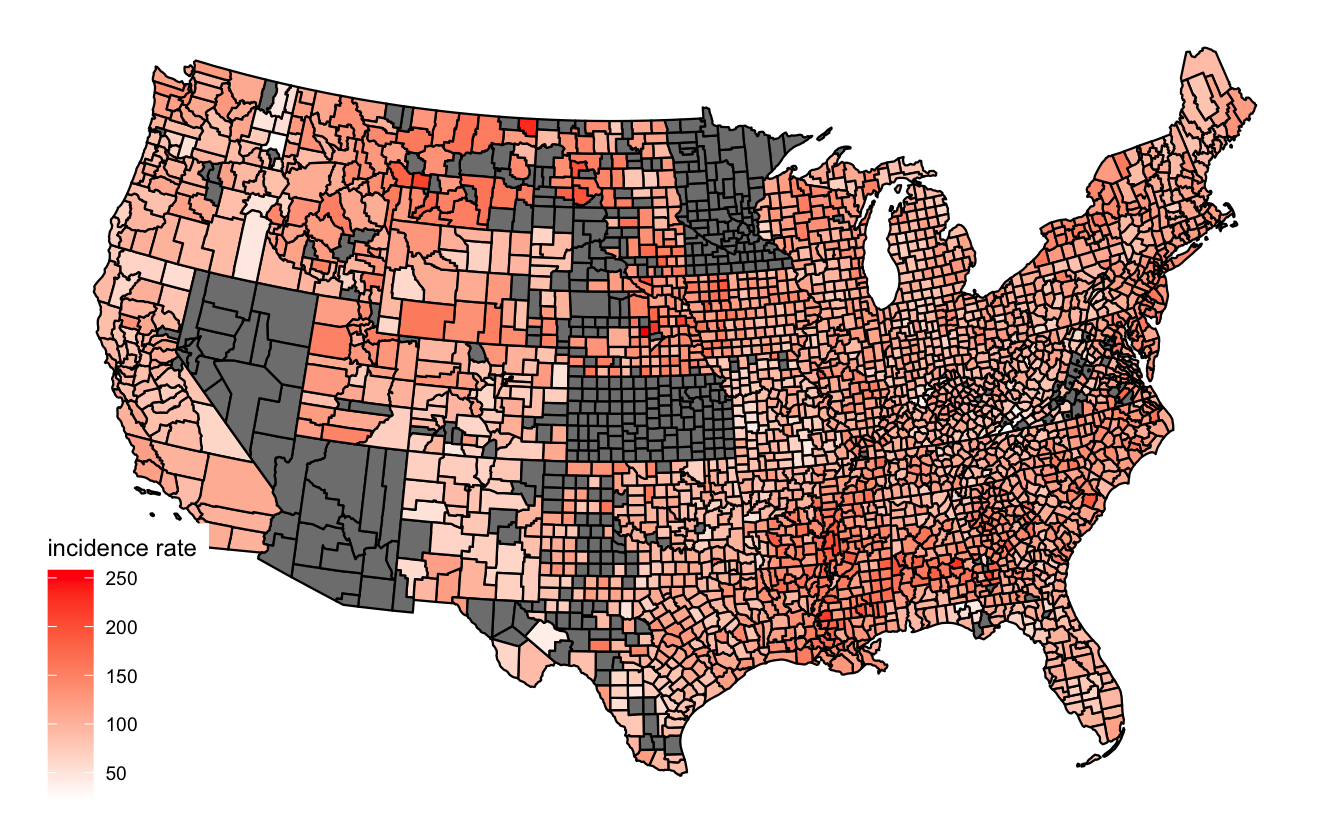}
        \caption{Prostate Cancer ($n=2623$)} 
        \label{}
    \end{subfigure}
    \hfill
    \begin{subfigure}[tb]{0.45\textwidth}
        \centering
        \includegraphics[width = \textwidth]{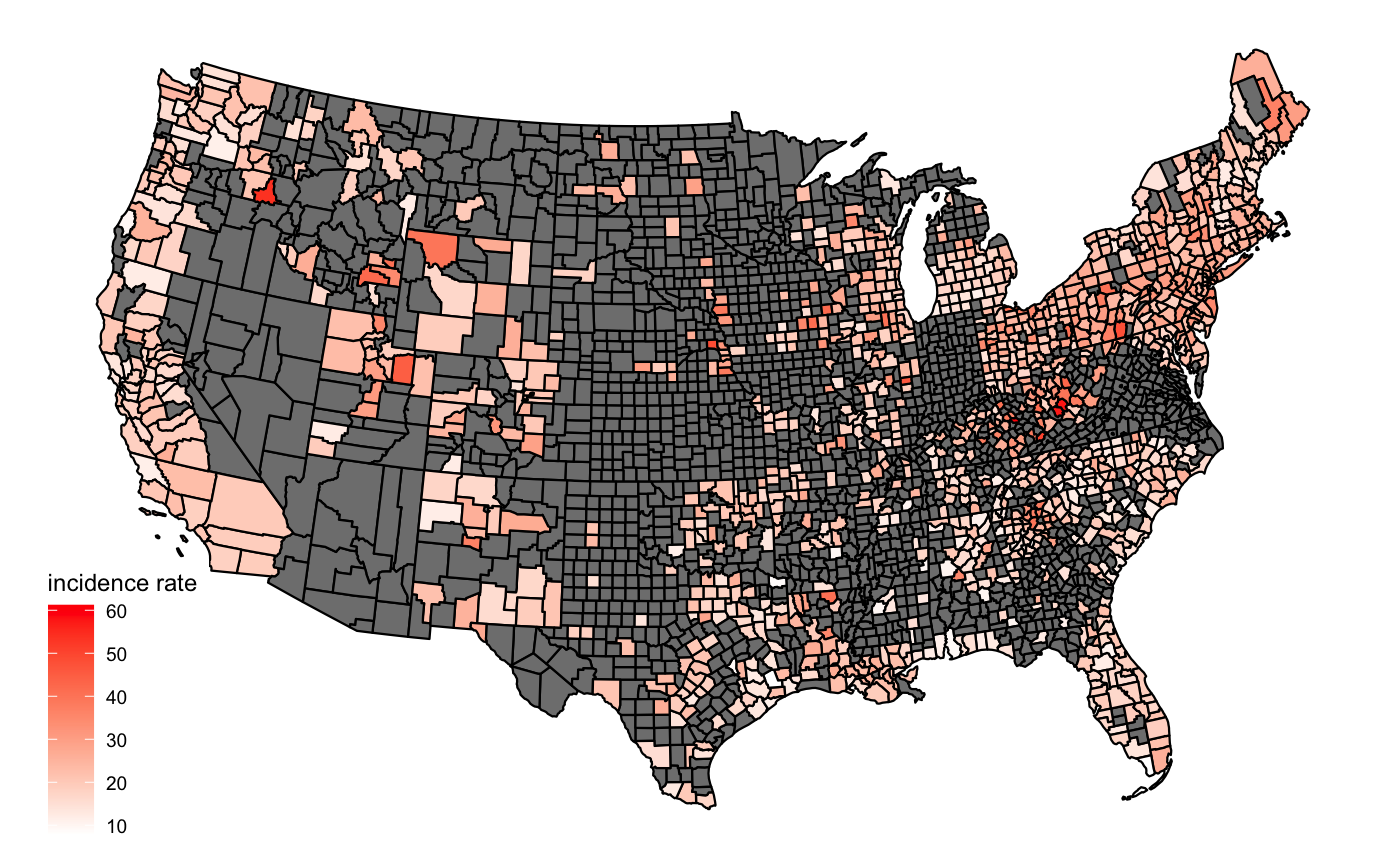}
        \caption{Thyroid Cancer ($n=1191$)} 
        \label{}
    \end{subfigure}
    \caption{U.S. Map of Cancer Incidence Rate for Hormonally Associated Cancers}{Supplementary Figure 4 shows maps of cancer incidence rate by county for four of the hormonally associated cancers.}
    \label{fig:hormon_cancer_ci_map}
\end{figure}

\clearpage
\begin{figure}[htb]
    \centering
    \begin{subfigure}[tb]{0.45\textwidth}
        \centering
        \includegraphics[width = \textwidth]{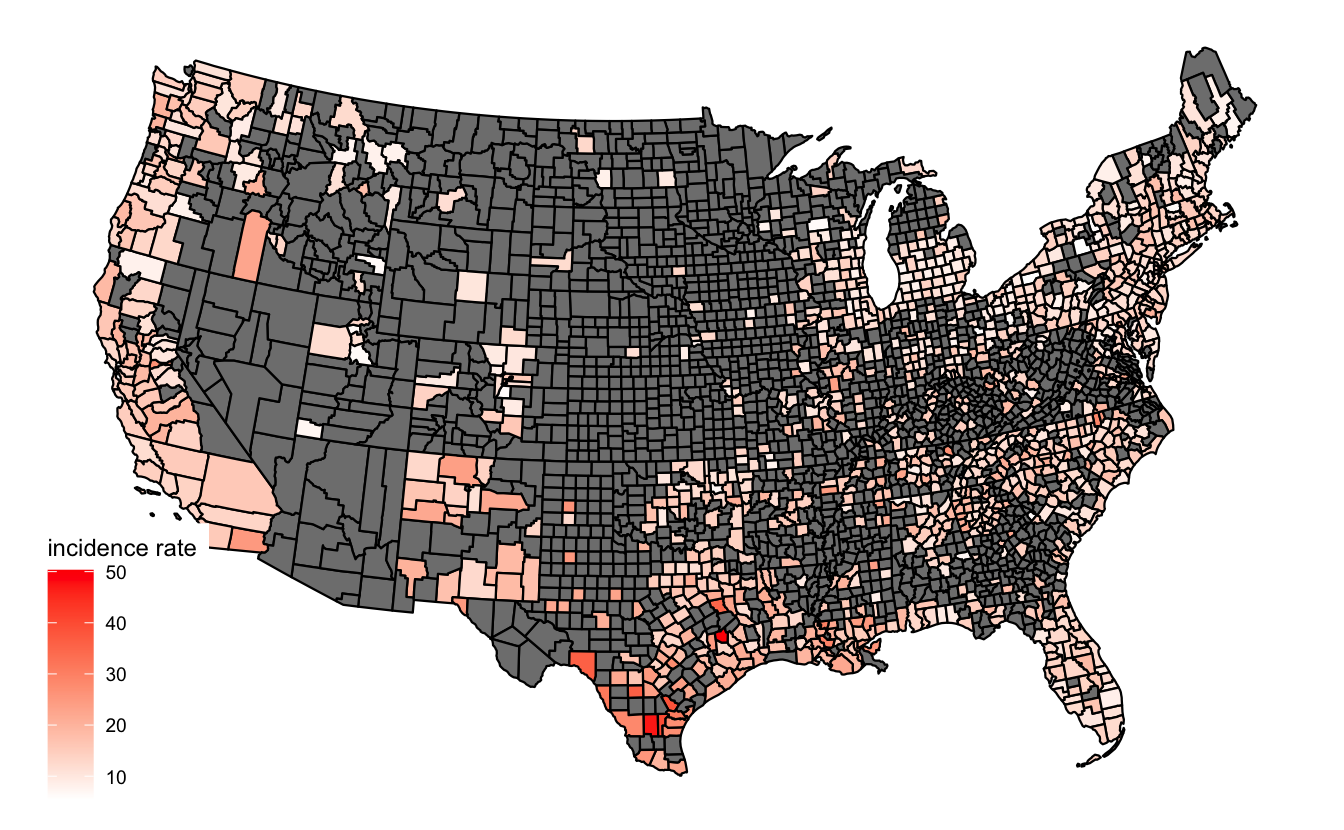}
        \caption{Liver \& Bile Duct Cancer ($n=1117$)} %
        \label{}
    \end{subfigure}
    \hfill
    \begin{subfigure}[tb]{0.45\textwidth}
        \centering
        \includegraphics[width = \textwidth]{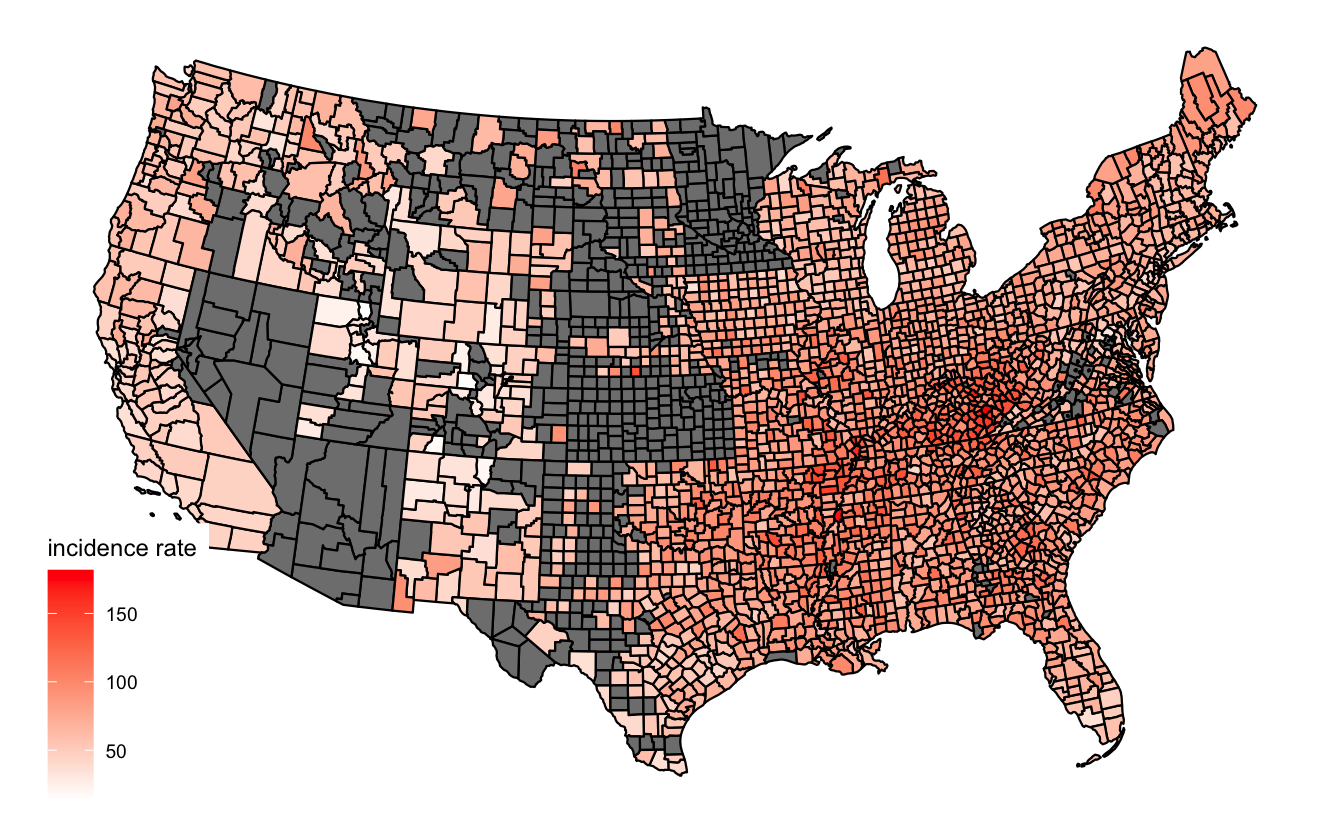}
        \caption{Lung \& Bronchus Cancer ($n=2441$)} %
        \label{}
    \end{subfigure} \\
    \begin{subfigure}[tb]{0.45\textwidth}
        \centering
        \includegraphics[width = \textwidth]{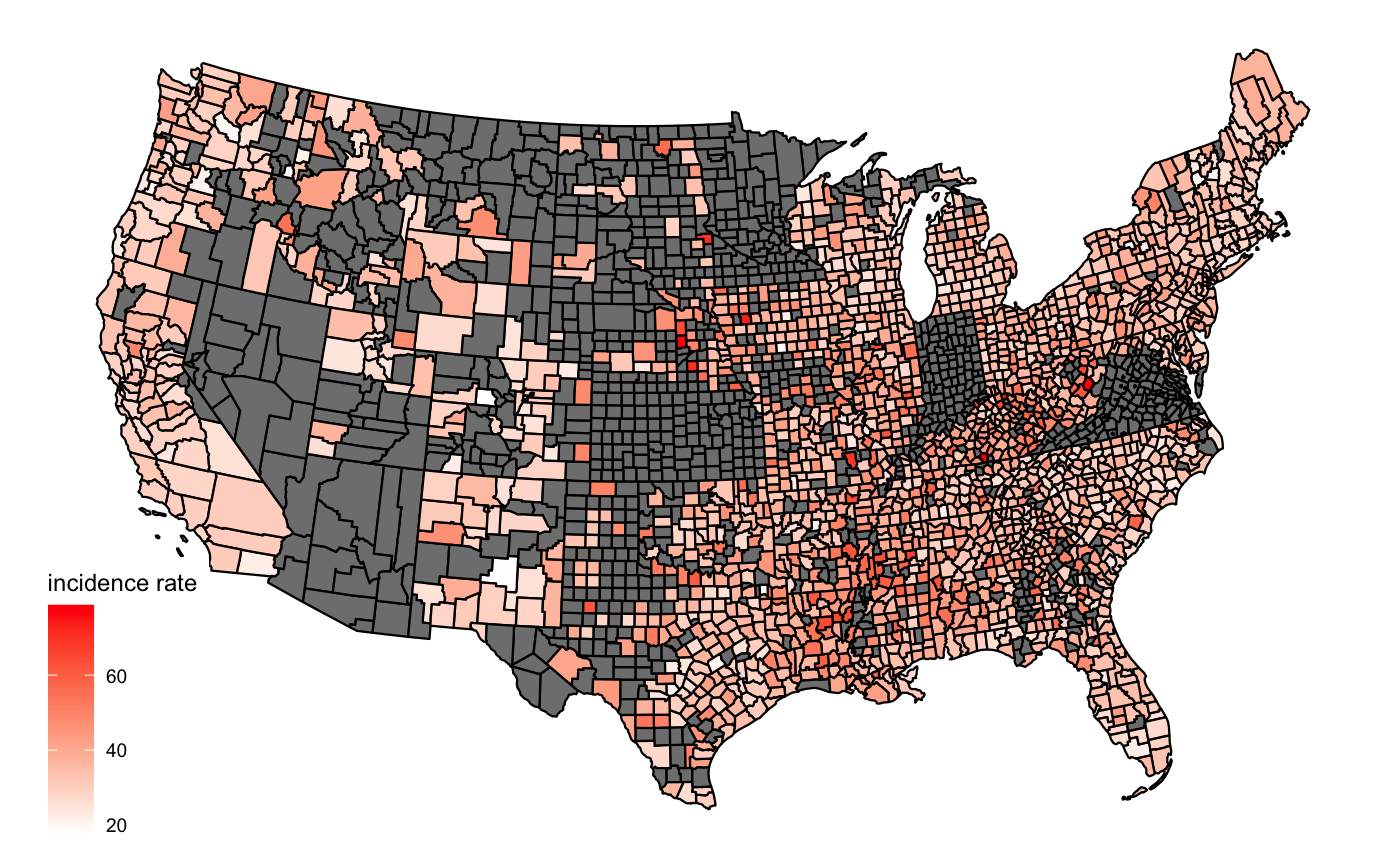}
        \caption{Colon \& Rectum Cancer ($n=1955$)} %
        \label{}
    \end{subfigure}
    \hfill
    \begin{subfigure}[tb]{0.45\textwidth}
        \centering
        \includegraphics[width = \textwidth]{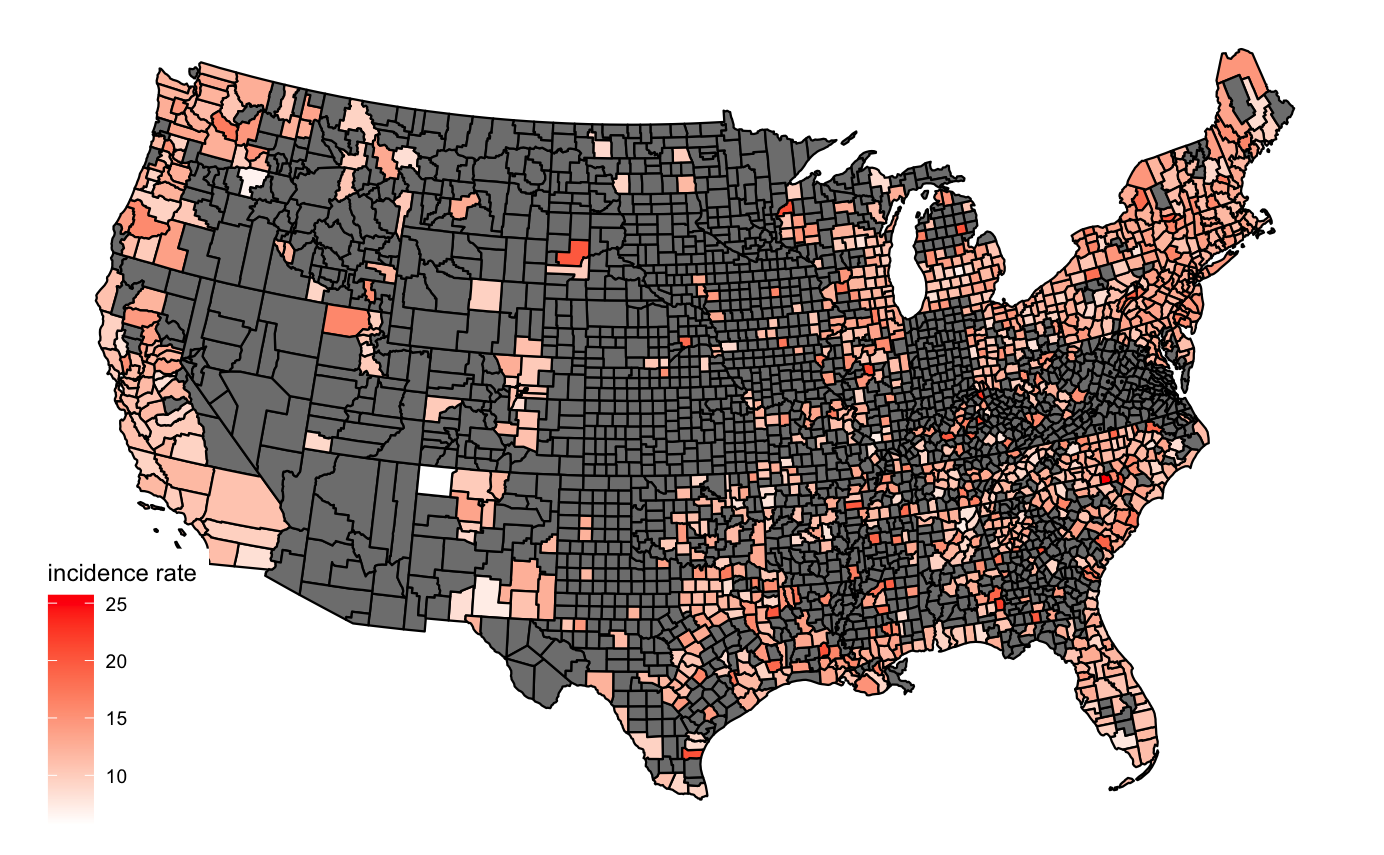}
        \caption{Pancreas Cancer ($n=1191$)} %
        \label{}
    \end{subfigure}
    \caption{U.S. Map of Cancer Incidence Rate for Four Most Prevalent Cancers}{Supplementary Figure 5 shows maps of cancer incidence rate by county for most prevalent cancer.}
    \label{fig:oth_cancer_ci_map}
\end{figure}

\clearpage
\begin{figure}[htb]
    \centering
    \includegraphics[width = 0.8\textwidth]{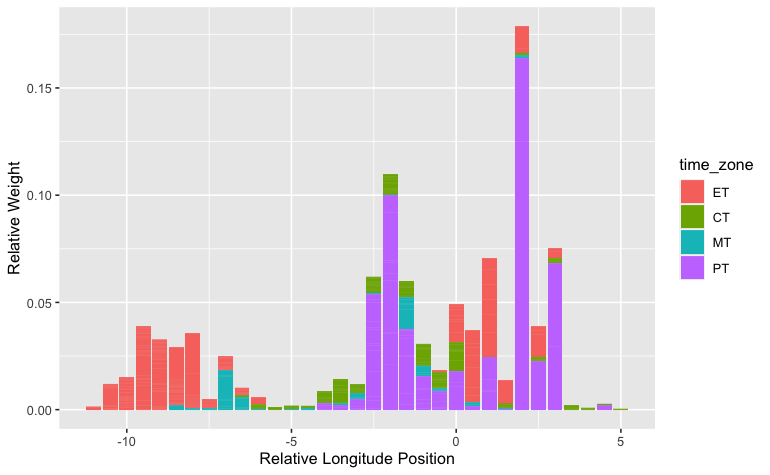}
    \caption{Relative Weights of 607 Counties
studied by Gu et al. (2017)}{Supplementary Figure 6 shows the relative weights of 607 counties
studied by Gu et al. (2017)}
    \label{fig:rel_weights}
\end{figure}

\clearpage
\begin{figure}[htb]
    \centering
    \includegraphics[width=0.7\textwidth]{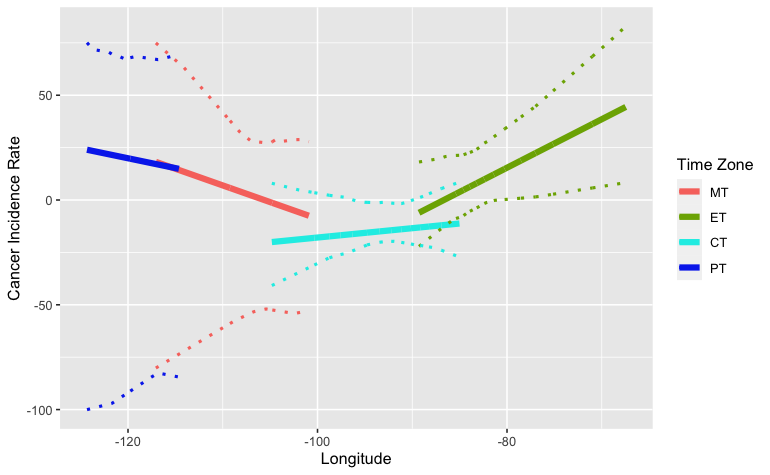}
    \caption{Output of Natural Splines for Total Cancer Incidence with Linear Approximation}{Supplementary Figure 7 shows the output of natural splines conducted on incidence by longitude and time zone for total cancer 
with 95\% bootstrap confidence band
using linear approximation ($n = 2853$)}
    \label{fig:linear_boot}
\end{figure}

\clearpage
\begin{table}[!htbp] \centering 
  \caption{Reported Coefficients of Relative Position for Hormonally Associated Cancers (with 95\% Confidence Interval)} 
  \label{linmod_relpos_hormon_cancer} 
\begin{adjustbox}{width = \textwidth}
{\begin{tabular}{@{\extracolsep{5pt}}lD{.}{.}{-3} D{.}{.}{-3} D{.}{.}{-3} D{.}{.}{-3} } 
\\[-1.8ex]\hline 
\hline \\[-1.8ex] 
 & \multicolumn{4}{c}{\textit{Dependent variable:}} \\ 
\cline{2-5} 
\\[-1.8ex] & \multicolumn{4}{c}{Cancer Incidence Rate} \\ 
 & \multicolumn{1}{c}{Breast} & \multicolumn{1}{c}{Ovary} & \multicolumn{1}{c}{Prostate} & \multicolumn{1}{c}{Thyroid} \\ 
\\[-1.8ex] & \multicolumn{1}{c}{(1)} & \multicolumn{1}{c}{(2)} & \multicolumn{1}{c}{(3)} & \multicolumn{1}{c}{(4)}\\ 
\hline \\[-1.8ex] 
 Relative Position & -0.285^{***} & -0.006 & 0.387^{***} & 0.123^{**} \\ 
  & \multicolumn{1}{c}{(-0.469$, $-0.101)} & \multicolumn{1}{c}{(-0.058$, $0.046)} & \multicolumn{1}{c}{(0.123$, $0.650)} & \multicolumn{1}{c}{(0.007$, $0.238)} \\ 
  & & & & \\ 
 Latitude & 0.514^{***} & -0.026 & 0.700^{***} & 0.090 \\ 
  & \multicolumn{1}{c}{(0.165$, $0.862)} & \multicolumn{1}{c}{(-0.113$, $0.060)} & \multicolumn{1}{c}{(0.198$, $1.203)} & \multicolumn{1}{c}{(-0.106$, $0.286)} \\ 
  & & & & \\ 
 High School & 45.179^{***} & 0.020 & 37.997^{**} & 21.493^{***} \\ 
  & \multicolumn{1}{c}{(20.773$, $69.584)} & \multicolumn{1}{c}{(-6.808$, $6.848)} & \multicolumn{1}{c}{(2.708$, $73.286)} & \multicolumn{1}{c}{(5.815$, $37.171)} \\ 
  & & & & \\ 
 Some College & 57.917^{***} & -0.461 & 86.469^{***} & 7.868 \\ 
  & \multicolumn{1}{c}{(37.167$, $78.666)} & \multicolumn{1}{c}{(-6.082$, $5.160)} & \multicolumn{1}{c}{(56.234$, $116.704)} & \multicolumn{1}{c}{(-5.232$, $20.967)} \\ 
  & & & & \\ 
 College and Above & 82.081^{***} & -2.537 & 71.577^{***} & 5.127 \\ 
  & \multicolumn{1}{c}{(60.290$, $103.872)} & \multicolumn{1}{c}{(-8.185$, $3.111)} & \multicolumn{1}{c}{(39.726$, $103.428)} & \multicolumn{1}{c}{(-8.184$, $18.438)} \\ 
  & & & & \\ 
 Elevation & -0.006^{***} & -0.0001 & -0.005^{***} & 0.001^{*} \\ 
  & \multicolumn{1}{c}{(-0.009$, $-0.003)} & \multicolumn{1}{c}{(-0.001$, $0.001)} & \multicolumn{1}{c}{(-0.009$, $-0.002)} & \multicolumn{1}{c}{(-0.0002$, $0.003)} \\ 
  & & & & \\ 
 Medical Doctor pc & 4.735 & -20.891 & -322.516 & -198.871^{**} \\ 
  & \multicolumn{1}{c}{(-296.566$, $306.035)} & \multicolumn{1}{c}{(-88.867$, $47.085)} & \multicolumn{1}{c}{(-773.165$, $128.133)} & \multicolumn{1}{c}{(-356.383$, $-41.359)} \\ 
  & & & & \\ 
 Median Income & 0.00004^{**} & -0.00000 & 0.00004 & 0.00000 \\ 
  & \multicolumn{1}{c}{(0.00000$, $0.0001)} & \multicolumn{1}{c}{(-0.00001$, $0.00000)} & \multicolumn{1}{c}{(-0.00001$, $0.0001)} & \multicolumn{1}{c}{(-0.00001$, $0.00002)} \\ 
  & & & & \\ 
 Obesity Rate & 22.247^{***} & -7.477^{***} & 15.190 & -8.401^{*} \\ 
  & \multicolumn{1}{c}{(8.124$, $36.369)} & \multicolumn{1}{c}{(-11.351$, $-3.603)} & \multicolumn{1}{c}{(-5.156$, $35.537)} & \multicolumn{1}{c}{(-17.135$, $0.333)} \\ 
  & & & & \\ 
 Smoking Rate & -105.899^{***} & 2.813 & -104.505^{***} & -32.751^{***} \\ 
  & \multicolumn{1}{c}{(-144.468$, $-67.330)} & \multicolumn{1}{c}{(-6.480$, $12.106)} & \multicolumn{1}{c}{(-161.113$, $-47.896)} & \multicolumn{1}{c}{(-54.674$, $-10.829)} \\ 
  & & & & \\ 
 PM2.5 (air pollution) & 0.138 & 0.110^{***} & 0.107 & 0.143^{*} \\ 
  & \multicolumn{1}{c}{(-0.155$, $0.431)} & \multicolumn{1}{c}{(0.042$, $0.178)} & \multicolumn{1}{c}{(-0.319$, $0.533)} & \multicolumn{1}{c}{(-0.016$, $0.301)} \\ 
  & & & & \\ 
 Water Violation & 1.408^{***} & 0.042 & 2.400^{***} & -0.139 \\ 
  & \multicolumn{1}{c}{(0.522$, $2.294)} & \multicolumn{1}{c}{(-0.163$, $0.246)} & \multicolumn{1}{c}{(1.119$, $3.681)} & \multicolumn{1}{c}{(-0.621$, $0.342)} \\ 
  & & & & \\ 
 Race (White) & -77.824^{**} & 9.754 & 209.432^{***} & -20.454 \\ 
  & \multicolumn{1}{c}{(-146.452$, $-9.195)} & \multicolumn{1}{c}{(-6.952$, $26.460)} & \multicolumn{1}{c}{(110.637$, $308.228)} & \multicolumn{1}{c}{(-58.808$, $17.900)} \\ 
  & & & & \\ 
 Race (Black) & -66.783^{*} & 9.113 & 288.464^{***} & -29.607 \\ 
  & \multicolumn{1}{c}{(-135.908$, $2.343)} & \multicolumn{1}{c}{(-7.731$, $25.958)} & \multicolumn{1}{c}{(188.715$, $388.212)} & \multicolumn{1}{c}{(-68.183$, $8.969)} \\ 
  & & & & \\ 
 Race (Native) & -105.620^{***} & 2.509 & 209.595^{***} & -20.723 \\ 
  & \multicolumn{1}{c}{(-178.470$, $-32.770)} & \multicolumn{1}{c}{(-15.283$, $20.300)} & \multicolumn{1}{c}{(104.050$, $315.139)} & \multicolumn{1}{c}{(-61.807$, $20.360)} \\ 
  & & & & \\ 
 Race (Asian) & -129.830^{***} & 10.906 & 157.325^{***} & -13.694 \\ 
  & \multicolumn{1}{c}{(-203.036$, $-56.624)} & \multicolumn{1}{c}{(-6.911$, $28.723)} & \multicolumn{1}{c}{(51.945$, $262.706)} & \multicolumn{1}{c}{(-54.732$, $27.343)} \\ 
  & & & & \\ 
 Race (Hispanic) & -89.638^{**} & 8.552 & 219.558^{***} & -13.322 \\ 
  & \multicolumn{1}{c}{(-158.906$, $-20.371)} & \multicolumn{1}{c}{(-8.331$, $25.435)} & \multicolumn{1}{c}{(119.763$, $319.353)} & \multicolumn{1}{c}{(-52.248$, $25.604)} \\ 
  & & & & \\ 
\hline \\[-1.8ex] 
Observations & \multicolumn{1}{c}{2,615} & \multicolumn{1}{c}{890} & \multicolumn{1}{c}{2,623} & \multicolumn{1}{c}{1,191} \\ 
R$^{2}$ & \multicolumn{1}{c}{0.534} & \multicolumn{1}{c}{0.276} & \multicolumn{1}{c}{0.573} & \multicolumn{1}{c}{0.504} \\ 
Adjusted R$^{2}$ & \multicolumn{1}{c}{0.523} & \multicolumn{1}{c}{0.226} & \multicolumn{1}{c}{0.563} & \multicolumn{1}{c}{0.478} \\ 
Residual Std. Error & \multicolumn{1}{c}{0.388 (df = 2554)} & \multicolumn{1}{c}{0.286 (df = 831)} & \multicolumn{1}{c}{0.612 (df = 2562)} & \multicolumn{1}{c}{0.475 (df = 1132)} \\ 
F Statistic & \multicolumn{1}{c}{48.760$^{***}$ (df = 60; 2554)} & \multicolumn{1}{c}{5.473$^{***}$ (df = 58; 831)} & \multicolumn{1}{c}{57.412$^{***}$ (df = 60; 2562)} & \multicolumn{1}{c}{19.825$^{***}$ (df = 58; 1132)} \\ 
\hline 
\hline \\[-1.8ex] 
\textit{Note:}  & \multicolumn{4}{r}{$^{*}$p$<$0.1; $^{**}$p$<$0.05; $^{***}$p$<$0.01} \\ 
\end{tabular} }
\end{adjustbox}
\end{table}

\clearpage
\begin{table}[!htbp] \centering 
  \caption{Reported Coefficients of Relative Position for Other Selected Cancers (with 95\% Confidence Interval)} 
  \label{linmod_relpos_subcancer} 
\begin{adjustbox}{width = \textwidth}{
\begin{tabular}{@{\extracolsep{5pt}}lD{.}{.}{-3} D{.}{.}{-3} D{.}{.}{-3} D{.}{.}{-3} D{.}{.}{-3} } 
\\[-1.8ex]\hline 
\hline \\[-1.8ex] 
 & \multicolumn{5}{c}{\textit{Dependent variable:}} \\ 
\cline{2-6} 
\\[-1.8ex] & \multicolumn{5}{c}{Cancer Incidence Rate} \\ 
 & \multicolumn{1}{c}{Colon \& Rectum} & \multicolumn{1}{c}{Liver \& Bile Duct} & \multicolumn{1}{c}{Liver \& Bile Duct (w/o MST)} & \multicolumn{1}{c}{Lung \& Bronchus} & \multicolumn{1}{c}{Pancreas} \\ 
\\[-1.8ex] & \multicolumn{1}{c}{(1)} & \multicolumn{1}{c}{(2)} & \multicolumn{1}{c}{(3)} & \multicolumn{1}{c}{(4)} & \multicolumn{1}{c}{(5)}\\ 
\hline \\[-1.8ex] 
 Relative Position & -0.011 & -0.084^{**} & -0.102^{***} & 0.140^{*} & 0.008 \\ 
  & \multicolumn{1}{c}{(-0.098$, $0.077)} & \multicolumn{1}{c}{(-0.158$, $-0.010)} & \multicolumn{1}{c}{(-0.179$, $-0.025)} & \multicolumn{1}{c}{(-0.023$, $0.302)} & \multicolumn{1}{c}{(-0.038$, $0.055)} \\ 
  & & & & & \\ 
 Latitude & 0.203^{***} & 0.071 & 0.033 & -0.025 & 0.132^{***} \\ 
  & \multicolumn{1}{c}{(0.053$, $0.353)} & \multicolumn{1}{c}{(-0.067$, $0.208)} & \multicolumn{1}{c}{(-0.110$, $0.175)} & \multicolumn{1}{c}{(-0.306$, $0.257)} & \multicolumn{1}{c}{(0.054$, $0.210)} \\ 
  & & & & & \\ 
 High School & 21.027^{***} & -5.935 & -8.192 & 16.051 & 9.797^{***} \\ 
  & \multicolumn{1}{c}{(9.925$, $32.128)} & \multicolumn{1}{c}{(-16.751$, $4.881)} & \multicolumn{1}{c}{(-19.274$, $2.890)} & \multicolumn{1}{c}{(-4.794$, $36.895)} & \multicolumn{1}{c}{(3.674$, $15.919)} \\ 
  & & & & & \\ 
 Some College & -1.367 & -1.847 & -3.303 & 27.186^{***} & 7.266^{***} \\ 
  & \multicolumn{1}{c}{(-10.666$, $7.932)} & \multicolumn{1}{c}{(-10.838$, $7.145)} & \multicolumn{1}{c}{(-12.572$, $5.966)} & \multicolumn{1}{c}{(9.601$, $44.770)} & \multicolumn{1}{c}{(2.119$, $12.413)} \\ 
  & & & & & \\ 
 College and Above & -4.064 & -13.262^{***} & -16.222^{***} & -22.630^{**} & 6.594^{**} \\ 
  & \multicolumn{1}{c}{(-13.668$, $5.539)} & \multicolumn{1}{c}{(-22.371$, $-4.154)} & \multicolumn{1}{c}{(-25.542$, $-6.903)} & \multicolumn{1}{c}{(-40.985$, $-4.274)} & \multicolumn{1}{c}{(1.381$, $11.807)} \\ 
  & & & & & \\ 
 Elevation & -0.001^{*} & -0.001^{**} & -0.002^{***} & -0.009^{***} & -0.0005 \\ 
  & \multicolumn{1}{c}{(-0.002$, $0.0002)} & \multicolumn{1}{c}{(-0.002$, $-0.0002)} & \multicolumn{1}{c}{(-0.003$, $-0.001)} & \multicolumn{1}{c}{(-0.011$, $-0.007)} & \multicolumn{1}{c}{(-0.001$, $0.0002)} \\ 
  & & & & & \\ 
 Medical Doctor pc & 15.124 & 430.093^{***} & 422.909^{***} & 958.870^{***} & 73.399^{**} \\ 
  & \multicolumn{1}{c}{(-104.990$, $135.238)} & \multicolumn{1}{c}{(318.346$, $541.841)} & \multicolumn{1}{c}{(309.565$, $536.252)} & \multicolumn{1}{c}{(711.023$, $1,206.717)} & \multicolumn{1}{c}{(7.949$, $138.849)} \\ 
  & & & & & \\ 
 Median Income & 0.00000 & 0.00000 & 0.00000 & 0.00001 & -0.00000 \\ 
  & \multicolumn{1}{c}{(-0.00001$, $0.00002)} & \multicolumn{1}{c}{(-0.00001$, $0.00002)} & \multicolumn{1}{c}{(-0.00001$, $0.00002)} & \multicolumn{1}{c}{(-0.00002$, $0.00004)} & \multicolumn{1}{c}{(-0.00001$, $0.00000)} \\ 
  & & & & & \\ 
 Obesity Rate & -3.134 & -4.150 & -5.287^{*} & 30.056^{***} & -1.452 \\ 
  & \multicolumn{1}{c}{(-9.439$, $3.171)} & \multicolumn{1}{c}{(-10.068$, $1.768)} & \multicolumn{1}{c}{(-11.323$, $0.750)} & \multicolumn{1}{c}{(18.115$, $41.998)} & \multicolumn{1}{c}{(-4.903$, $1.998)} \\ 
  & & & & & \\ 
 Smoking Rate & 10.528 & 39.361^{***} & 35.483^{***} & 226.326^{***} & 8.799^{**} \\ 
  & \multicolumn{1}{c}{(-5.842$, $26.897)} & \multicolumn{1}{c}{(24.681$, $54.041)} & \multicolumn{1}{c}{(20.469$, $50.498)} & \multicolumn{1}{c}{(194.047$, $258.606)} & \multicolumn{1}{c}{(0.133$, $17.465)} \\ 
  & & & & & \\ 
 PM2.5 (air pollution) & 0.004 & 0.018 & 0.013 & 0.346^{***} & 0.008 \\ 
  & \multicolumn{1}{c}{(-0.118$, $0.127)} & \multicolumn{1}{c}{(-0.092$, $0.127)} & \multicolumn{1}{c}{(-0.099$, $0.125)} & \multicolumn{1}{c}{(0.115$, $0.577)} & \multicolumn{1}{c}{(-0.055$, $0.071)} \\ 
  & & & & & \\ 
 Water Violation & -0.032 & -0.344^{**} & -0.320^{*} & 0.399 & 0.043 \\ 
  & \multicolumn{1}{c}{(-0.399$, $0.335)} & \multicolumn{1}{c}{(-0.670$, $-0.018)} & \multicolumn{1}{c}{(-0.650$, $0.011)} & \multicolumn{1}{c}{(-0.325$, $1.122)} & \multicolumn{1}{c}{(-0.149$, $0.235)} \\ 
  & & & & & \\ 
 Race (White) & -28.977^{*} & -51.620^{***} & -56.709^{***} & -241.304^{***} & -3.439 \\ 
  & \multicolumn{1}{c}{(-58.255$, $0.300)} & \multicolumn{1}{c}{(-78.410$, $-24.830)} & \multicolumn{1}{c}{(-84.721$, $-28.697)} & \multicolumn{1}{c}{(-298.072$, $-184.536)} & \multicolumn{1}{c}{(-18.929$, $12.051)} \\ 
  & & & & & \\ 
 Race (Black) & -23.891 & -45.369^{***} & -50.392^{***} & -246.785^{***} & 0.912 \\ 
  & \multicolumn{1}{c}{(-53.412$, $5.630)} & \multicolumn{1}{c}{(-72.414$, $-18.324)} & \multicolumn{1}{c}{(-78.647$, $-22.137)} & \multicolumn{1}{c}{(-304.098$, $-189.472)} & \multicolumn{1}{c}{(-14.714$, $16.538)} \\ 
  & & & & & \\ 
 Race (Native) & -30.286^{*} & -59.876^{***} & -68.771^{***} & -311.201^{***} & -10.055 \\ 
  & \multicolumn{1}{c}{(-61.974$, $1.403)} & \multicolumn{1}{c}{(-89.073$, $-30.678)} & \multicolumn{1}{c}{(-101.536$, $-36.006)} & \multicolumn{1}{c}{(-371.743$, $-250.659)} & \multicolumn{1}{c}{(-27.049$, $6.940)} \\ 
  & & & & & \\ 
 Race (Asian) & -32.624^{**} & -39.608^{***} & -45.250^{***} & -224.304^{***} & -3.502 \\ 
  & \multicolumn{1}{c}{(-63.812$, $-1.436)} & \multicolumn{1}{c}{(-68.239$, $-10.976)} & \multicolumn{1}{c}{(-75.110$, $-15.389)} & \multicolumn{1}{c}{(-284.614$, $-163.994)} & \multicolumn{1}{c}{(-20.001$, $12.996)} \\ 
  & & & & & \\ 
 Race (Hispanic) & -30.277^{**} & -43.964^{***} & -50.357^{***} & -259.359^{***} & -2.417 \\ 
  & \multicolumn{1}{c}{(-59.853$, $-0.701)} & \multicolumn{1}{c}{(-71.080$, $-16.847)} & \multicolumn{1}{c}{(-78.675$, $-22.040)} & \multicolumn{1}{c}{(-316.784$, $-201.934)} & \multicolumn{1}{c}{(-18.072$, $13.237)} \\ 
  & & & & & \\ 
\hline \\[-1.8ex] 
Observations & \multicolumn{1}{c}{1,955} & \multicolumn{1}{c}{1,117} & \multicolumn{1}{c}{1,063} & \multicolumn{1}{c}{2,441} & \multicolumn{1}{c}{1,090} \\ 
R$^{2}$ & \multicolumn{1}{c}{0.418} & \multicolumn{1}{c}{0.556} & \multicolumn{1}{c}{0.541} & \multicolumn{1}{c}{0.823} & \multicolumn{1}{c}{0.334} \\ 
Adjusted R$^{2}$ & \multicolumn{1}{c}{0.401} & \multicolumn{1}{c}{0.530} & \multicolumn{1}{c}{0.516} & \multicolumn{1}{c}{0.818} & \multicolumn{1}{c}{0.296} \\ 
Residual Std. Error & \multicolumn{1}{c}{0.318 (df = 1896)} & \multicolumn{1}{c}{0.404 (df = 1056)} & \multicolumn{1}{c}{0.404 (df = 1007)} & \multicolumn{1}{c}{0.431 (df = 2380)} & \multicolumn{1}{c}{0.265 (df = 1031)} \\ 
F Statistic & \multicolumn{1}{c}{23.511$^{***}$ (df = 58; 1896)} & \multicolumn{1}{c}{21.996$^{***}$ (df = 60; 1056)} & \multicolumn{1}{c}{21.576$^{***}$ (df = 55; 1007)} & \multicolumn{1}{c}{184.027$^{***}$ (df = 60; 2380)} & \multicolumn{1}{c}{8.909$^{***}$ (df = 58; 1031)} \\ 
\hline 
\hline \\[-1.8ex] 
\textit{Note:}  & \multicolumn{5}{r}{$^{*}$p$<$0.1; $^{**}$p$<$0.05; $^{***}$p$<$0.01} \\ 
\end{tabular} 
}\end{adjustbox}
\end{table}

\clearpage
\begin{table}[htb] \centering
    \caption{Types of Cancer Included in Total Cancer Incidence Rate (19 total)}
    \label{cancertype}
\begin{tabular}{c}
\hline
Cancer Type            \\ \hline
Bladder                \\
Brain \& ONS           \\
Breast                 \\
Cervix                 \\
Colon \& rectum        \\
Esophagus              \\
Kidney \& renal pelvis \\
Leukemia               \\
Liver \& bile duct     \\
Lung \& bronchus       \\
Melanoma of the skin   \\
Non-hodgkin lymphoma   \\
Oral cavity \& pharynx \\
Ovary                  \\
Pancreas               \\
Prostate               \\
Stomach                \\
Thyroid                \\
Uterus                 \\ \hline
\end{tabular}
\end{table}

\clearpage
\begin{table}[htb] \centering
    \caption{Counties and Cities combined in Virginia}
    \label{virginia_append}
\begin{tabular}{c}
\hline
County + City \\ \hline
Albemarle + Charlottesville              \\
Alleghany + Covington                    \\
Augusta, Staunton + Waynesboro           \\
Campbell + Lynchburg                     \\
Carroll + Galax                          \\
Dinwiddie, Colonial Heights + Petersburg \\
Fairfax, Fairfax City + Falls Church     \\
Frederick + Winchester                   \\
Greensville + Emporia                    \\
Henry + Martinsville                     \\
James City + Williamsburg                \\
Montgomery + Radford                     \\
Pittsylvania + Danville                  \\
Prince George + Hopewell                 \\
Prince William, Manassas + Manassas Park \\
Roanoke + Salem                          \\
Rockbridge, Buena Vista + Lexington      \\
Rockingham + Harrisonburg                \\
Southampton + Franklin                   \\
Spotsylvania + Fredericksburg            \\
Washington + Bristol                     \\
Wise + Norton                            \\
York + Poquoson                          \\ \hline
\end{tabular}
\end{table}

\clearpage
\begin{table}[!htb]
\centering
    \caption{\commentnj{Summary Table of Results Obtained by Gu et al. and Present Paper}}
    \label{sum_gu_results}
\begin{adjustbox}{width = \textwidth}{
\begin{tabular}{ccc}
\hline
\multicolumn{1}{|c|}{Topic}                                                                                                                                                                     & \multicolumn{1}{c|}{Findings of Gu et al.}                                                                          & \multicolumn{1}{c|}{Findings of this paper}                                                                                                                                                                                 \\ \hline
\multicolumn{1}{|c|}{\begin{tabular}[c]{@{}c@{}}Number of observations for \\ total cancer rate (counties)\end{tabular}}                                                                    & \multicolumn{1}{c|}{607}                                                                                            & \multicolumn{1}{c|}{2853}                                                                                                                                                                                                   \\ \hline
\multicolumn{1}{|c|}{Race and ethnicity studied}                                                                                                                                                & \multicolumn{1}{c|}{White}                                                                                          & \multicolumn{1}{c|}{\begin{tabular}[c]{@{}c@{}}White, black, American Indian / AK native, \\ Asian / Pacific Islander, Hispanic (any race)\end{tabular}}                                                                    \\ \hline
\multicolumn{1}{|c|}{Total cancer (inconsistent)}                                                                                                                                           & \multicolumn{1}{c|}{RR* = 1.029 (significant**)}                                                                    & \multicolumn{1}{c|}{RR\_eqv*** = 2.000}                                                                                                                                                                                     \\ \hline
\multicolumn{1}{|c|}{Breast cancer (consistent)}                                                                                                                                                & \multicolumn{1}{c|}{RR = 1.074}                                                                                     & \multicolumn{1}{c|}{RR\_eqv = 1.425 (significant)}                                                                                                                                                                          \\ \hline
\multicolumn{1}{|c|}{Ovary cancer}                                                                                                                                                              & \multicolumn{1}{c|}{—}                                                                                              & \multicolumn{1}{c|}{RR\_eqv = 0.030}                                                                                                                                                                                        \\ \hline
\multicolumn{1}{|c|}{Prostate cancer (inconsistent)}                                                                                                                                            & \multicolumn{1}{c|}{RR = 1.042 (significant)}                                                                       & \multicolumn{1}{c|}{RR\_eqv = -1.935 (significant)}                                                                                                                                                                         \\ \hline
\multicolumn{1}{|c|}{Thyroid cancer}                                                                                                                                                            & \multicolumn{1}{c|}{RR = 1.042}                                                                                     & \multicolumn{1}{c|}{RR\_eqv = -0.615}                                                                                                                                                                                       \\ \hline
\multicolumn{1}{|c|}{Liver and bile duct cancer (consistent)}                                                                                                                                   & \multicolumn{1}{c|}{RR = 1.110 (significant)}                                                                       & \multicolumn{1}{c|}{RR\_eqv = 0.510 (significant)}                                                                                                                                                                          \\ \hline
\multicolumn{1}{|c|}{Lung and bronchus cancer}                                                                                                                                                  & \multicolumn{1}{c|}{RR = 1.002}                                                                                     & \multicolumn{1}{c|}{RR\_eqv = -0.700}                                                                                                                                                                                       \\ \hline
\multicolumn{1}{|c|}{Colon and rectum cancer}                                                                                                                                                   & \multicolumn{1}{c|}{RR = 1.023}                                                                                     & \multicolumn{1}{c|}{RR\_eqv = 0.055}                                                                                                                                                                                        \\ \hline
\multicolumn{1}{|c|}{Pancreas cancer}                                                                                                                                                           & \multicolumn{1}{c|}{RR = 1.041}                                                                                     & \multicolumn{1}{c|}{RR\_eqv = -0.040}                                                                                                                                                                                       \\ \hline
\multicolumn{3}{l}{\begin{tabular}[c]{@{}l@{}}* Response variable used by Gu et al. is covariate-adjusted rate ratio (RR) per 5 degrees difference in longitude, \\ which is equivalent to five times the regression coefficient of our model in magnitude. \\ Furthermore, they interpret their RR as “moving from the east to the west” while we interpret our coefficient as \\ moving from the west to the east. Therefore, we also flip the sign of their RR in order to make the comparison \\ with our estimated coefficients.\end{tabular}} \\
\multicolumn{3}{l}{** All significance represents the 0.01 significance level.}                                                                                                                                                                                                                                                                                                                                                                                                                                                                     \\
\multicolumn{3}{l}{\begin{tabular}[c]{@{}l@{}}*** RR\_eqv is derived from the coefficients in our model through the progress described above (namely RR\_eqv \\ = -coef*5).\end{tabular}}                                                                                                                                                                                                                                                                                                                                                          
\end{tabular} }
\end{adjustbox}
\end{table}

%% file: mainSRDD.bbl
\begin{thebibliography}{10}

\bibitem{gu2017longitude}
F.~Gu, S.~Xu, S.~S. Devesa, F.~Zhang, E.~B. Klerman, B.~I. Graubard, and N.~E.
  Caporaso, ``Longitude position in a time zone and cancer risk in the united
  states,'' {\em Cancer Epidemiology and Prevention Biomarkers}, vol.~26,
  no.~8, pp.~1306--1311, 2017.

\bibitem{smith2016spring}
A.~C. Smith, ``Spring forward at your own risk: Daylight saving time and fatal
  vehicle crashes,'' {\em American Economic Journal: Applied Economics},
  vol.~8, no.~2, pp.~65--91, 2016.

\bibitem{lahti2006transition}
T.~A. Lahti, S.~Lepp{\"a}m{\"a}ki, J.~L{\"o}nnqvist, and T.~Partonen,
  ``Transition to daylight saving time reduces sleep duration plus sleep
  efficiency of the deprived sleep,'' {\em Neuroscience letters}, vol.~406,
  no.~3, pp.~174--177, 2006.

\bibitem{barnes2009changing}
C.~M. Barnes and D.~T. Wagner, ``Changing to daylight saving time cuts into
  sleep and increases workplace injuries.,'' {\em Journal of applied
  psychology}, vol.~94, no.~5, p.~1305, 2009.

\bibitem{bunnings2021spring}
C.~B{\"u}nnings and V.~Schiele, ``Spring forward, don't fall back: the effect
  of daylight saving time on road safety,'' {\em Review of economics and
  statistics}, vol.~103, no.~1, pp.~165--176, 2021.

\bibitem{carey2017impact}
R.~N. Carey and K.~M. Sarma, ``Impact of daylight saving time on road traffic
  collision risk: a systematic review,'' {\em BMJ open}, vol.~7, no.~6,
  p.~e014319, 2017.

\bibitem{rishi2020daylight}
M.~A. Rishi, O.~Ahmed, J.~H. Barrantes~Perez, M.~Berneking, J.~Dombrowsky,
  E.~E. Flynn-Evans, V.~Santiago, S.~S. Sullivan, R.~Upender, K.~Yuen, {\em
  et~al.}, ``Daylight saving time: an american academy of sleep medicine
  position statement,'' {\em Journal of clinical sleep medicine}, vol.~16,
  no.~10, pp.~1781--1784, 2020.

\bibitem{MeiraeCruz2019}
M.~M. e~Cruz, M.~Miyazawa, R.~Manfredini, D.~Cardinali, J.~Madrid, R.~Reiter,
  J.~Araujo, R.~Agostinho, and D.~Acu{\~{n}}a-Castroviejo, ``Impact of daylight
  saving time on circadian timing system: An expert statement,'' {\em European
  Journal of Internal Medicine}, vol.~60, pp.~1--3, Feb. 2019.

\bibitem{young2017circadian}
M.~E. Young, ``Circadian control of cardiac metabolism: Physiologic roles and
  pathologic implications,'' {\em Methodist DeBakey cardiovascular journal},
  vol.~13, no.~1, p.~15, 2017.

\bibitem{heboyan2019effects}
V.~Heboyan, S.~Stevens, and W.~V. McCall, ``Effects of seasonality and daylight
  savings time on emergency department visits for mental health disorders,''
  {\em The American journal of emergency medicine}, vol.~37, no.~8,
  pp.~1476--1481, 2019.

\bibitem{borisenkov2011latitude}
M.~F. Borisenkov, ``Latitude of residence and position in time zone are
  predictors of cancer incidence, cancer mortality, and life expectancy at
  birth,'' {\em Chronobiology international}, vol.~28, no.~2, pp.~155--162,
  2011.

\bibitem{giuntella2019sunset}
O.~Giuntella and F.~Mazzonna, ``Sunset time and the economic effects of social
  jetlag: evidence from us time zone borders,'' {\em Journal of health
  economics}, vol.~65, pp.~210--226, 2019.

\bibitem{open-elevation}
J.~R. Lourenço, ``Open elevation.''
  \url{https://github.com/Jorl17/open-elevation}, 2021.

\end{thebibliography}
